\documentclass[sigconf, nonacm]{acmart}
\usepackage{svg}
\usepackage{amsfonts}

\usepackage[ruled,vlined]{algorithm2e}
\SetAlgoSkip{} 
\IncMargin{1em} 

\usepackage{algpseudocode}
\usepackage{graphicx}
\usepackage{textcomp}
\usepackage{epsfig}
\usepackage{tcolorbox}
\usepackage{graphicx,amsmath,amssymb,subfigure,stfloats, url}
\usepackage{multirow}
\usepackage{listings}
\usepackage{booktabs}
\usepackage{balance}
\usepackage{soul}
\usepackage{hyperref}
\usepackage[normalem]{ulem}
 \usepackage[T1]{fontenc}
\usepackage{aecompl}
\usepackage{mdframed}
\usepackage{booktabs} 
\usepackage{multirow} 
\usepackage{tabularx} 
\usepackage{listings}
\usepackage{array}
\usepackage{caption}
\usepackage{subcaption}
\usepackage{makecell}
\usepackage{listings}
\usepackage{colortbl}

\newcommand\vldbdoi{XX.XX/XXX.XX}

\newcommand{\eat}[1]{}
\newcommand{\stitle}[1]{\noindent{\bf #1}}

\newcommand{\eg}{\emph{e.g.,}\xspace}

\begin{document}
\title{ER-RAG: Enhance RAG with ER-Based Unified Modeling of Heterogeneous Data Sources}
\author{Yikuan Xia$^{*1}$, Jiazun Chen$^{*1}$, Yirui Zhan$^1$, Suifeng Zhao$^1$, Weipeng Jiang$^2$,\\ Chaorui Zhang$^2$, Wei Han$^2$, Bo Bai$^{\dagger2}$, Jun Gao$^{\dagger1}$}
 \thanks{$^{*}$ Equal Contribution. $^{\dagger}$Corresponding Authors.}
\affiliation{$^1$Key Laboratory of High Confidence Software Technologies, CS, Peking University, Beijing, China}
\affiliation{$^2$Theory Lab, Central Research Institute, 2012 Labs, Huawei Technologies Co., Ltd, Huawei Research, China}
\email{wfl00014@pku.edu.cn, {chenjiazun,zhanyirui}@stu.pku.edu.cn,zsf030826@163.com}
\email{{jiangweipeng,zhang.chaorui,harvey.hanwei,baibo8}@huawei.com, gaojun@pku.edu.cn}
\begin{abstract}

Large language models (LLMs) have demonstrated significant potential in performing question-answering (QA) tasks. Retrieval-augmented generation (RAG) enhances the reliability and precision of LLMs through the integration of external evidence beyond their training data. The supporting evidence is retrieved from external sources, including web pages, databases, and knowledge graphs, among other structured and unstructured data sources.  While effective in many scenarios, current RAG methods require dedicated retrieval agents for individual data sources to optimize retrieval quality. Such an agent-specific strategy presents significant challenges in low-resource or black-box environments and complicates operations when evidence is fragmented across sources.

To address the above limitations, we propose ER-RAG, a framework designed to enable RAG systems to integrate information from multiple heterogeneous data sources. ER-RAG adopts the generalized Entity-Relationship (ER) model to represent evidence across diverse physical data forms, inspired by the observation of restrictive access patterns in the RAG context. The framework encapsulates different data sources with ER-based unified application programming interfaces (APIs), allowing entity retrieval and relationship querying through standardized $GET$ and $JOIN$ operations. Moreover, ER-RAG employs a two-stage generation process for API chain construction: in the first stage, a direct preference optimization-tuned module identifies optimal sources, and in the second stage, a dedicated module generates API chains based on the schema of the selected sources. This unified access pattern enables ER-RAG to accumulate extensive training samples for module fine-tuning while facilitating seamless integration across diverse data sources. A previous version of ER-RAG, tailored exclusively to the MockAPI and the provided web pages, won first place in all three tracks of the 2024 KDDCup CRAG Challenge. Experimental results demonstrate that ER-RAG, utilizing an 8B LLM backbone, achieves performance comparable to commercial RAG pipelines built on state-of-the-art LLM backbones. ER-RAG also surpasses other hybrid source competitors by 3.1\% in LLM score and accelerates the retrieval process by 5.5X. \eat{Experimental results show that ER-RAG achieves comparative performance to the source-specific and commercial RAGs, and significantly improves the accuracy and efficiency on multi-source QA tasks.}
\end{abstract}

\maketitle


\section{Introduction}

Large language models (LLMs), typically pre-trained on extensive data, have led to significant advancements in various natural language processing tasks, particularly in question-answering (QA) systems~\cite{LLMfewshot}. Retrieval-augmented generation (RAG) has emerged as an effective approach, requiring minimal to no additional training. RAG is particularly valuable when the desired knowledge lies outside the LLMs' training corpus~\cite{ragsurvey24,crag2024}. Specifically, RAG employs a retriever component to search for and extract relevant evidence from external data sources. The evidence is then integrated into the LLM's context window to enhance its generation capabilities.
\begin{figure} [t]
    \centering
    \includegraphics[width=\columnwidth]{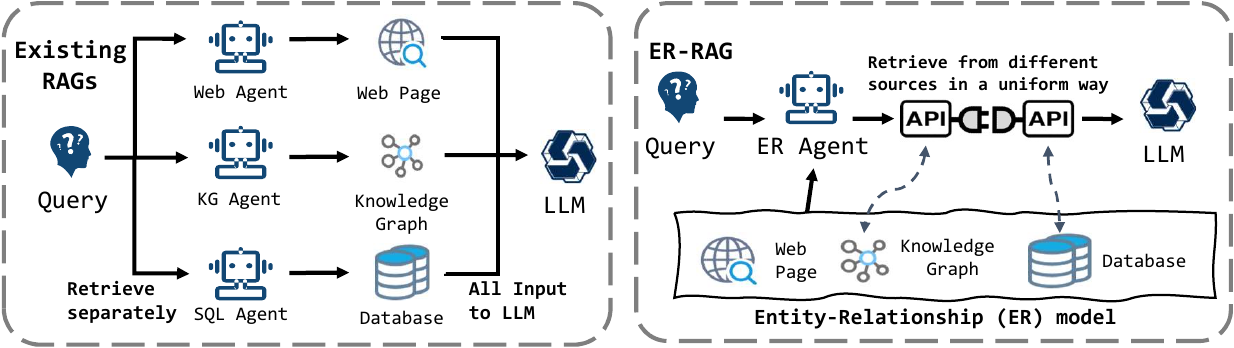}
    \vspace{-0.5cm}
    \caption{Existing RAGs vs. ER-RAG: Traditional RAG creates separate agents per source, while ER-RAG uses ER modeling to unify integration via ER API.}
    \label{fig:difference}
     \vspace{-0.5cm}
\end{figure}

Real-world RAG applications encounter diverse external data sources and their various combinations \cite{graphrag, hybridrag}. Web pages, including news articles and blog posts \cite{rag_chunk_23}, serve as sources of dynamic, real-time information. Databases provide structured data repositories, particularly for organizational records such as inventory and customer information~\cite{tag}.  Knowledge graphs facilitate extracting semantically rich and interconnected knowledge, with Wikipedia serving as a prominent example~\cite{tog1}. Additionally, closed-source data sources, \eg web service for querying stock information, typically return structured JSON-formatted information in response to entity-specific queries \cite{json}. While specific data sources excel at certain queries, multi-source QA tasks (\eg merging corporate and open-web data) inherently require cross-source evidence.

\eat{For some complex tasks like multiple-hop question-answers, the partial evidence may be in one data source, while the resting part may be in other sources.}

Existing RAG systems adopt strategies tailored to the characteristics of different data sources, as illustrated in Figure \ref{fig:difference}. For web pages, the content is pre-processed and chunked into smaller pieces, and the chunks are matched with the query using vector-based semantic similarity to identify relevant information~\cite{rag_chunk_23}. In some cases, more advanced approaches, such as graph-based relation modeling~\cite{tog1}, are employed to capture complex relationships within the chunks. For databases, existing methods either convert database content into a vector database for semantic matching~\cite{sigtabret} or generate SQL queries to extract the required information~\cite{tog1}. For knowledge graphs, existing methods often use vector-based retrieval to extract relevant nodes or subgraphs~\cite{kgvector}, and some other systems deploy agents that traverse nodes and edges to gather information~\cite{tog1, tog2}. These diverse techniques enable RAG systems to adapt to the unique requirements of each data source.

\eat{. Certain methods generate SPARQL queries to retrieve structured knowledge~\cite{SPARQL},}

While effective in certain scenarios, existing RAG systems face significant challenges in handling heterogeneous sources. First, \textbf{the separate heterogeneous data retrievers increase  more training and maintenance difficulty.} As access methods vary among multiple sources, distinct training sets must be prepared for each retriever, leading to limited training data for individual sources under fixed computational and maintenance costs. These resource constraints further limit the adaptation of the RAG system to new data sources.

\eat{While effective in certain scenarios, existing RAG systems face significant challenges in multi-source context.}

Second, \textbf{information interaction between different data sources remains challenging}. For queries requiring evidence from multiple data sources, \eg $S_1$ and $S_2$, simply combining their respective retrieval results may lead to irrelevant information being retrieved and erroneous answers being generated. An alternative approach involves developing a complex answering agent that generates multiple queries on one source (\eg $S_1$) based on query results from another source (\eg $S_2$), which substantially increases computational and operational costs.

\eat{An alternative is a complex answering agent generating a long chain of thought interacting with the data sources for multiple times~\cite{}, which may lead to a significantly higher cost.}

Third, \textbf{choosing a proper data source combination to enhance both efficiency and accuracy in source selection remains a critical challenge.} Since evidence relevant to a given query may exist across different sources, including the LLM itself, which incurs no external retrieval cost, or external sources with varying access costs (\eg open search engines being more costly than internal databases), it is crucial to carefully select the most appropriate sources for retrieval. While prior work \cite{selfrag} has explored the use of self-reflection to determine whether retrieval is necessary, the challenge of balancing efficiency and accuracy when retrieving from multiple sources remains under-explored.

We observe that RAG excels at handling knowledge-intensive tasks ~\cite{rag_icml20}, where the evidence required in RAG can be reduced to the location of entities and their relationships~\cite{tag}, \eg fetching the age of a person from Wikipedia, fetching the student number of a particular student's name from a database, and retrieving the entity (person) that has a certain relationship (author) with the second entity (book). With full consideration of such characteristics, we leverage the classical ER concept model \cite{ermodel}, and propose a unified ER-based API over heterogeneous data sources, regardless of their physical format. In practice, both open-box sources such as DB, Wikipedia, and WEB, and closed-box sources can fit into the unified API.  Such unified APIs enable training a single retrieval module to perform operations across the selected sources using a single training dataset. The key contributions to the above challenges are listed as follows:

\eat{or retrieving a character's friend's names from a novel document. The question may also be related to the relationship between entities. For example, the first entity's attribute $a_1$ is equal to the second entity's attribute $a_2$, or }

\hangafter 1
\hangindent 2em
1. ER-RAG standardizes APIs for multiple data sources, whose generalization is backed up by the ER model~\cite{ermodel}. The APIs consist of a $GET$ and a $JOIN$ component. The $GET$ component describes querying entities from a certain data source, filtering some of the entities by conditions, and getting their desired attributes.  Provided with a consistent interface, LLMs can perform reasoning more effectively. 

\hangafter 1
\hangindent 2em
2. ER-RAG introduces two-stage generation for the chain of APIs, in which the proper sources are first selected by a module fine-tuned by DPO to boost efficiency and accuracy, and the chains of APIs are then generated by another module with the schema of the selected source via the prompts. Such a separation makes the module fine-tuning easier, and lowers the context length of the generation module with a more concise schema. ER-RAG also adopts a post-processing module to manipulate collected data and maintain system expressiveness.

\hangafter 1
\hangindent 2em
3. Our experimental results indicate that ER-RAG achieves performance comparable to source-specific RAG systems while significantly enhancing the accuracy and efficiency of multi-source QA tasks. Specifically, ER-RAG demonstrates a 3.1\% improvement in LLM scores over other hybrid source competitors and accelerates the retrieval process by 5.5X. ER-RAG also show robust generalization capabilities, effectively adapting to datasets that differ from its original training data. These results highlight ER-RAG's ability to bridge the gap between heterogeneous data sources and advanced reasoning capabilities in LLMs, showcasing its potential as a unified framework for integrating diverse data sources.

The paper proceeds as follows: Section~\ref{sec-related} reviews related work, Section~ \ref{sec-api} details unified APIs, Section~\ref{sec-errag} outlines ER-RAG components, Section~ \ref{sec-exp} presents experimental results, and Section~\ref{sec-conclusion} concludes with future directions.

\eat{
The paper proceeds as follows: We review the related work in Section~\ref{sec-related}. Then, we describe the unified access APIs and their implementations in Section~\ref{sec-api}, and present the key components in ER-RAG in Section~\ref{sec-errag}. Section~\ref{sec-exp} reports the experimental results. We conclude the paper and foresee the future work in Section~\ref{sec-conclusion}.
}

\begin{figure*} [t]
    \centering
    \includegraphics[width=1\textwidth]{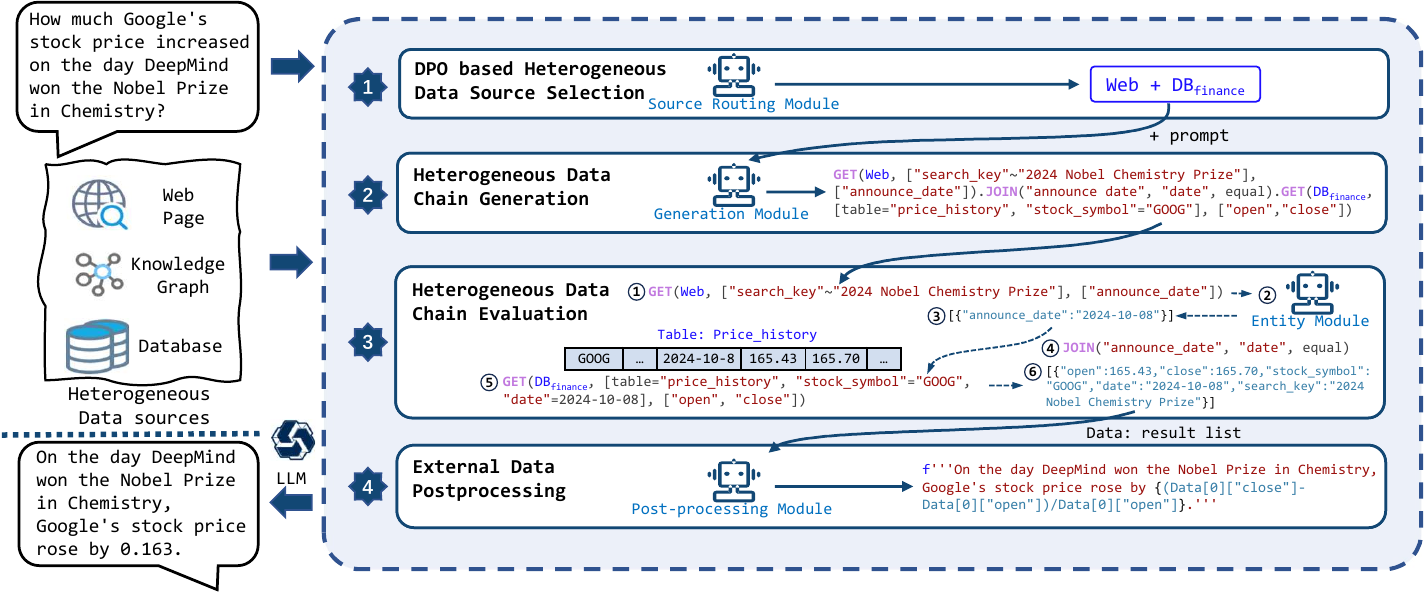}
    \vspace{-0.3cm}
    \caption{\eat{Illustration of ER-RAG inference pipeline. In the first step, a module is introduced to choose the data source that provides sufficient information and low retrieval costs.  Both DB\_Finance and Wikipedia datasets are required to retrieve the stock price and Nobel Prize announcement date.  In the second step, an API chain is generated given a prompt describing the data schema of the chosen data sources. We first query the date and use it to retrieve the stock price data. In the third step, the generated API chain is executed to get the retrieval results. The retrieval result is also formatted in JSON, containing attributes extracted from heterogeneous data sources.  In the final step, a post-processing module is introduced to process the retrieved information into natural language f-string formats with executable Python expressions in case post-computation is needed to answer the query.}
    ER-RAG inference pipeline: (1) A source selection module identifies optimal data sources optimizing information sufficiency and retrieval cost. Both DB\_Finance and WEB datasets are required to retrieve the stock price and Nobel Prize announcement date.  (2) An API chain is generated based on selected sources' schema, first querying temporal data (Nobel Prize date) to enable stock price lookup. (3) The API chain executes to return JSON-formatted results from heterogeneous sources. (4) A post-processor converts outputs to natural language f-string templates with executable Python expressions.}
    \label{fig:task2framework}
     \vspace{-0.3cm}
\end{figure*}

\section{Related work}
\label{sec-related}
This section reviews cross-source RAG, application-specific source selection, LLM fine-tuning strategies, and unified heterogeneous data modeling.

\subsection{RAG in Different Kinds of Sources}

\stitle{RAG over Text Sources.} HTML and text sources are the major external data sources for current RAG systems, and there is a significant body of work in this area. Usually, the content of the HTML pages is decomposed into chunks \cite{rag_chunk_23}, each of which contains a sequential piece of documents. Subsequently, the relevant chunks can be located using a combined metric with sparse and dense retrieval. The sparse retrieval mainly captures explicit matching of the keywords in the query and chunks using TF-IDF \cite{tfidf} or BM25~\cite{bm25}. The dense retrieval \cite{chen2024bge} focuses on the implicit semantic matching between the query and chunks if both are converted into embeddings using a pre-trained language model like BERT~\cite{bert}. The existing retrieval strategies are viewed as the implementations of the following unified ER-based APIs. For example, we can construct an API with the output attribute named "chunk", and we can follow the metrics to locate the most relevant chunks.

\stitle{RAG on Knowledge Graph.} Knowledge graphs~\cite{kg_Dong2023}, constructed on the open corpus or internal resources, can provide high-quality supplementary context to enhance the performance of LLMs. When adopted by RAG, the user's query has to be converted into the query over the knowledge graph using the template \cite{templateparsekg} or flexible semantic parser \cite{sematicparsekg}. Since the generated paths may not align with those in existing knowledge graphs, Think-on-Graph (TOG) \cite{tog1} introduces an LLM to select among candidate paths for the RAG. Existing methods serve as knowledge graph access implementations seamlessly integrated into ER-RAG.

\stitle{RAG on Relational Database.} Relational databases, such as storing an organization's internal data, can provide high-quality references for RAG systems when handling questions in natural language.  Natural language to SQL (NL2SQL/TEXT2SQL) has recently become a hot research topic \cite{NLP2SQLDawn, gao2024text}, involving prompt design, fine-tuned LLMs, or pre-trained language models combined with small task-specific models. Another related task, table retrieval, can be viewed as a sub-task of NL2SQL, which involves retrieving the relevant single table \cite{sigtabret} or multiple tables with join conditions \cite{multabret} by computing the similarities between queries and tables. In our context, however, we focus on how to integrate information between the database and other data sources, rather than extracting from the database alone. Among these works on relational databases, the task most closely related to ours is Table-Augmented Generation (TAG)~\cite{tag}, which features LLM executions in SQL to allow semantic understanding in using relational databases as a source of RAG. ER-RAG shows its difference from TAG due to more heterogeneous data sources, stronger semantic understanding using information from other sources, and fine-tuned LLMs for the source selection and generation of the chain of APIs.

\eat{, like the graph-aware enhancements or skeleton-aware enhancements}

\stitle{RAG on Hybrid Sources.} Recently, RAG systems have started to incorporate hybrid sources for evidence collection, aiming to enhance the breadth and quality of the information used in response generation. For example, GraphRAG \cite{graphrag} constructs a knowledge graph from text sources and applies a clustering algorithm to the graph. The resulting clusters support answers to global queries. HybridRAG \cite{hybridrag} combines GraphRAG \cite{graphrag} with a basic text-based RAG system, enhancing performance by utilizing retrieved data from both sources. Think-on-graph 2.0 \cite{tog2} enhances the selection of navigation paths in the knowledge graph guided by both LLM and collected documents. We can see the need for hybrid sources in RAG, especially when knowledge graphs are constructed on the fly from text sources. However, unlike ER-RAG, all existing works rely on specific methods to support access to relational databases, knowledge graphs, and text sources separately.

\subsection{Source Selection and LLM Routing}

Given a query, the source selection has been investigated in different application domains via a task-specific metric considering the cost, quality, and reliability of the data sources. For example, a cost-efficiency metric is proposed in data integration~\cite{less_Dong2012}, and a trustworthiness score is computed~\cite{Trust_source2015} in the web source answering. In the context of RAG, as the user's query is flexible, the selection of proper sources requires a deep understanding of the features of the query and the capability of the data sources, which can be facilitated by the fine-tuned LLM in our proposed method.

Another related problem is to route queries to the proper LLM to achieve a trade-off between performance and cost, as there are a large number of LLMs with different capabilities and expenses~\cite{routebench}. The current routers usually learn routing policies to maximize the pre-defined objectives based on the understanding of the user's query and LLMs. Among them, RoutLLM~\cite{routellm} casts the problem as a supervised classification problem with a point-wise loss, and LLM-BLENDER~\cite{routellmreward} introduces a reward model to evaluate the routing results with a pair-wise loss. ER-RAG can recommend multiple sources instead of a single source using a fine-tuned LLM, as the evidence needed may be located in different sources.

\subsection{Fine-Tuning of LLM}

It is a critical step to fine-tune pre-trained LLMs in handling specific tasks, and supervised fine-tuning (SFT) and reward modeling are two major fine-tuning strategies. SFT prepares an annotated dataset with explicit outputs to optimize the parameters in the LLM directly. Reward modeling provides fine-granularity preference control over tuned LLM by SFT, among which RLHF \cite{RLHF} utilizes human feedback to train a reward model that guides the actions of LLM. Representative reinforcement learning methods include proximal policy optimization (PPO)  \cite{ppo}, and direct preference optimization (DPO) \cite{dpo}. Specifically, DPO attempts to optimize the actions in each step without requiring explicit reward modeling. ER-RAG adopts the SFT in building the module in generating the chain of the API, and takes an SFT-then-DPO approach in fine-tuning the module in source selection to achieve the preference in combined metrics with cost and accuracy. 

Specifically, fine-tuning can be implemented in Low-Rank Adaptation (LoRA)~\cite{lora} to further enhance parameter efficiency and adaptability in fine-tuning. LoRA introduces trainable low-rank decomposition matrices into pre-trained LLMs, significantly reducing trainable parameters while maintaining task performance. This paper implements all tuned sub-modules via LoRA adapters, enabling low-cost switching and training.
\subsection{Heterogeneous Data Source Modeling}

The ER model is a classical generalized conceptual framework used to describe the key concepts to be managed in a database, including entities, attributes, and their relationships \cite{ermodel}.  The ER model can be viewed as the foundation of the design of physical models, including knowledge graphs, JSON, and relational databases. By fully considering the restricted access pattern in RAG, we leverage the idea of the ER model to define unified access APIs over all data sources to obtain the necessary evidence (entities and relationships).

There are also works to represent different data sources using one unified model. As the relational database schema is derived from ER modeling, relational tables can be reconstructed into a graph with foreign keys as the edges between table nodes. Previous works learn embeddings at the schema or instance level using graph neural networks to support various downstream tasks, such as recommendation \cite{atjnet}, entity resolution, and table understanding \cite{embdi,entitygraph}. As RAG access patterns are restricted, ER-RAG's unified APIs serve as basic graph query operations, with their composition forming a chain rather than a flexible graph pattern.

\section{Unified ER API Chain}
\label{sec-api}

In this section, we first define our unified ER API for knowledge retrieval from heterogeneous data sources. We then describe the concrete physical implementation of the defined logical API. Finally, we outline the greedy execution algorithm of the API chain.
\subsection{Unified API's Logical Definition}

Motivated by the ER model and entity-entry MockAPI in CRAG dataset~\cite{crag2024}, we design a unified ER extraction API to extract knowledge from different sources. First, we define two basic operations of the API, the $GET$ operation and the $JOIN$ operation.

\stitle{GET Operation.} The $GET$ operation corresponds to the process of retrieving specific entities and attributes from a data source, defined by a set of filtering criteria. In the context of the ER model, the $GET$ operation can be viewed as a selection process over an entity set $\{E\}$, where the selection condition is based on attribute values.  

Formally, let \( S \) represent a data source, and let \( A_1, A_2, \dots, A_n \) denote the attributes of \( \{E\} \). A $GET$ operation can be expressed as:
\begin{equation*}
 GET(S, \textit{Condition}, \{A_i\})
\end{equation*}
 where \( S \) is the target data source, \textit{Condition} is a logical expression involving one or more attributes of \( \{E\} \), and  \( \{A_i\} \) is the set of attributes to be retrieved. The result of a $GET$ operation is a projection of specific attributes of condition meeting entities $\{E'\} \subseteq \{E\}$.

\stitle{JOIN Operation.} The $JOIN$ operation is used to combine data from two entity sets based on two related attributes. In the ER model, this operation reflects the establishment of relationships between entities. A $JOIN$ operation integrates information by linking entities through shared attributes.  $JOIN$ operation takes two $GET$ outputs (entity subsets with associated attributes) as input. 

Formally, let \( S_1 \) and \( S_2 \) represent two data sources, and let \(\{A^1_i\} \) and \( \{A^2_i\} \) be attributes in \( S_1 \) and \( S_2 \), respectively. The output of two $GET$s is \( \{E_1\} \) and \( \{E_2\} \). The $JOIN$ operation can be expressed as:
\begin{equation*}
GET(S_1,...,\{A^1_i\}).JOIN(\textit{Condition}).GET(S_2,...,\{A^2_i\}).
\end{equation*}
 where $\textit{Condition}=(e_1.a_1\ \text{op}\ e_2.a_2)$ defines the relationship between the entities, and $e_1 \in \{E_1\}, e_2 \in \{E_2\}$. The used attributes have to be concluded in the extracted attributes, $a_1 \in \{A^1_i\}, a_2 \in \{A^2_i\}$. The output of the $JOIN$ operation is a subset $\{E_{12}'\} \subseteq \{E_1\} \times \{E_2\}$ (Cartesian product of the two sets $\{E_1\}$ and $\{E_2\}$), where the pairs in $\{E_{12}'\}$ satisfy the specified $JOIN$ condition.

\stitle{Chain of GET and JOIN.} Ideally, the $JOIN$ operation can be placed between $GET$ operations and form a graph of $GET$ operations, allowing complex relationships between data sources to be expressed and extracted. Considering that QA queries often do not involve such complex logic represented in graphs, we consider a chain of alternating $GET$ and $JOIN$ operations to simplify the generation and representation of these processes.  This chain represents a sequence of data retrieval and integration steps, where each $GET$ operation extracts specific entities and attributes, and each $JOIN$ operation combines these entities based on defined conditions. Formally, a chain of $GET$ and $JOIN$ operations can be expressed as:
\begin{align*}
&GET(S_1, \text{Condition}_1, \{A^1_i\}) .\text{JOIN}(\text{Condition}_2).GET(S_2,  \text{Condition}_3,  \{A^2_i\})  \\
&.\text{JOIN}(\text{Condition}_4).GET(S_3, \text{Condition}_5, \{A^3_i\}) \dots
\end{align*}

Such a chain alternates between $GET$ and $JOIN$ operations, building unified multi-source data views.
\subsection{Unified API's Physical Implementation}
This subsection details the API system's physical implementation and enumerates representative instances. Notably, the API execution has no limitations on data sources, so other sources can also be included as long as our defined APIs are implemented.

\stitle{Instances of GET APIs.} We list the implementations of $GET$ APIs for several common data sources here, including the DB, WIKI, and Other (WEB, Document) sources. 

\textbf{\textit{DB GET API.}} In the context of relational databases, the $GET$ operation can be directly translated into an SQL query. The $GET$ operation in a DB is formally expressed as:
$GET(DB_s, \textit{Condition}, \{A_i\})$,
where \( DB_s \) is the data source (typically a database containing multiple related tables), \textit{Condition} represents a logical expression involving one or more attributes in the database table schema, and \( \{A_i\} \) is the set of attributes to be retrieved. Specifically, in the \textit{Condition} statements, a certain table has to be specified for successful retrieval ($\textit{table = table\_name}$). We represent the rest of the conditions as $\textit{Condition}'$. In SQL, this operation can be represented as:
\vspace{-0.1cm}
\begin{equation*}
\textit{SELECT }  \{A_i\} \textit{ FROM } \textit{table\_name} \textit{ WHERE } \textit{Condition}';.
\end{equation*}
Here, \( \{A_i\} \) are the attributes of interest, and the $\textit{WHERE Condition}'$ specifies the filtering criteria for selecting the rows from the table \( S \). The \textit{Condition'} in the $GET$ operation often corresponds to SQL-like conditions, which can involve equality comparisons (\eg \( A_i = v \)), range queries (\eg \( A_i > v \)), pattern matching (\eg \text{LIKE},\textasciitilde), or logical combinations of these (\eg \( A_i = v_1 \text{ AND } B_i < v_2 \)). Specifically for the pattern matching case, fuzzy search is conducted by first attempting an exact match, followed by a case-insensitive partial match using ILIKE (PostgreSQL) for prefix or suffix completion. For queries on specific fields with a limited set of attribute values and the above search result is empty, embedding-based retrieval is utilized to identify the most relevant attribute.

\eat{Specifically for the pattern matching case, we build a BM25 \cite{bm25} index in the database for efficiently handling pattern matching queries, especially for frequently queried key fields like names of people or movies, as it improves both performance and relevance of search results. }

\textbf{\textit{{Knowledge Graph GET API.}}} In the context of Knowledge Graphs (KG), the $GET$ operation is used to retrieve specific attributes or facts about entities.  We demonstrate the KG $GET$ interface using Wikipedia:  $GET(WIKI,$ $\textit{Condition}, \{A_i\})$.

In the condition part, we use the same condition statements as the DB $GET$ API instance. A specific search\_key in the statement should be provided to identify the entity we needed. For example, if the \textit{Condition} contains search\_key= DeepMind, we use Wikipedia search tools to get the page of DeepMind for later processing. To deal with the ambiguity problem~\cite{sevgili2022neural}, we prompt an 8B LLM to select the most related entity given the query and disambiguation list. Furthermore, we distill a entity module (for entity disambiguation and later  attribute extraction) using GPT-4o outputs to enhance its cross-modal reasoning capabilities. Other condition statements listed in the DB GET API can also be incorporated. We additionally leverage it to verify whether the retrieved Wikipedia entity satisfies the specified conditions.

In the attribute part, if $\{A_i\}$ (\eg "parent company" of a company) exists directly in the structured infobox or table of the entity's page (\eg the Wikipedia page of DeepMind), it can be extracted directly using the structured data parser. However, if the attribute is not explicitly available in the structured data, or the attribute name can't be exactly matched but exists in the unstructured text of the page, LLM can be employed to extract the desired attribute. We prompt the LLM to process the text and identify the relevant information. Specifically, we also fine-tune the entity module with the output of GPT-4o to enhance its capability. By extracting structured data when available and unstructured data using LLM, this hybrid approach enables efficient and comprehensive retrieval of attributes in KGs, ensuring both precision and flexibility in handling diverse query types.

\textbf{\textit{Other GET API.}} Other GET APIs follow the same format as the other two instances. Other sources, mainly web pages extracted from search engine results or documents, follow the same text processing pipeline \cite{xia2024}. To process information from web pages, the HTML content is first parsed into structured documents. We use the BeautifulSoup library to extract the main body text while filtering out irrelevant elements. After extraction, or when dealing with plain text, we use TextSplitter from LangChain to divide the text into manageable chunks \cite{rag_chunk_23}, adhering to the predefined parent-child chunking strategy. These chunks are then fed into the retriever and reranker pipeline. The retriever identifies the most relevant child chunks based on their semantic similarity to the query, while the pairwise reranker model further refines the selection, boosting the retrieval performance. We use the retrieved chunk results as a default "chunk" attribute, and we also support the LLM attribute extraction module mentioned in the KG GET API part to extract attributes from these retrieved chunks. Notably, we implement a widely-used simple RAG pipeline for the web and text scenario, and other advanced RAG techniques, \eg query rewriting~\cite{gao2023precise}, or HTML parsing techniques, \eg extracting the table contents, can be applied to further boost the performance, which is perpendicular to our research direction.

\stitle{Instances of JOIN API.} A naive instance of the $JOIN$ operation can be described as follows: Suppose that we have two data sources \(S_1\) and \(S_2\), with attributes \(\{A^1_i\}\) and \(\{A^2_i\}\), respectively. The output of the first $GET$ operation is $E_1$, a set of entities with attributes.
Let the $JOIN$ \textit{Condition} be defined as \(e_1.a_1 \ \text{op} \ e_2.a_2\), where \(a_1 \in \{A^1_i\}\) and \(a_2 \in \{A^2_i\}\). For each entity \(e_1 \in E_1\), the $JOIN$ condition generates a filter for the second $GET$ operation, where the condition is concatenated with logical OR for all \(e_1 \in E_1\). Formally, the resulting $JOIN$ condition can be expressed as:
\vspace{-0.1cm}
\begin{equation*}
\textit{Condition} = \bigvee_{e_1 \in E_1} \left(e_2.a_2 \ \text{op} \ e_1.a_1\right),
\vspace{-0.1cm}    
\end{equation*}
where \(\bigvee\) denotes the logical OR operator. This means that for each entity \(e_1\) in the output of the first $GET$, a corresponding condition $e_2.a_2 \ \text{op} \ e_1.a_1$ is inserted into the second $GET$ operation. The $JOIN$ operation integrates the outputs \(E_1\) and \(E_2\) by filtering the Cartesian product \(E_1 \times E_2\) based on the defined condition.

Notably, for simple operands, \eg =,>,<, we can simply generate the conditions of the second $GET$. However, for semantic operands, like the parent company operand between DeepMind and Google, semantic understanding is required to parse such operands. Therefore, to handle the parsing of the semantic operands, we integrate the entity module (originally designed for the KG GET API)  in our $JOIN$ interface for KG and text. For example,  for a $JOIN$ interface $JOIN(content,\ parent\_company,\ search\_key)$, that queries the parent company of DeepMind. We take the content attribute of the first $GET$ of DeepMind's wiki page, which is a default attribute of all WIKI-related $GET$s, and prompt the entity module to extract a parent company entity from the content. The answer is filled into a condition related to the second $GET$'s $search\_key$, fetching the Google entity for further processing.

\eat{\stitle{Instances of Cardinality API.} In the context of interface implementation, it is often beneficial, though optional, to provide a cardinality estimate for each $GET$ instance. Cardinality refers to the number of records or entities that satisfy the conditions of a specific $GET$ operation. By estimating cardinality, we can optimize the execution order of multiple $GET$ and $JOIN$ operations, like SQL execution in the database \cite{card}, as operations with lower cardinality can be prioritized to reduce intermediate results and improve overall efficiency. For database (DB) APIs, cardinality can be estimated using metadata (\eg \text{min}, \text{max}, and \text{distinct\_count}) and the converted SQL query. Our system uses a simple estimation approach based on this metadata. For KG APIs and other APIs, if a \text{search\_key} is provided, the cardinality is estimated as \(1\), assuming the key uniquely identifies a single entity. If no \text{search\_key} is provided, the cardinality is estimated as \(\infty\), reflecting that we can't identify the entity based on the current interface.}

\subsection{Execution of Chain of APIs}

\eat{Interface Parsing}

The execution order of APIs has a significant impact on execution time, and optimizing the sequence can greatly improve efficiency. To determine the execution order of each $GET$ operation, we need to have access to the cardinality of each $GET$ operation. Cardinality refers to the number of records or entities that satisfy the conditions of a specific $GET$ operation. By estimating cardinality, we can optimize the execution order of multiple $GET$ and $JOIN$ operations, like SQL execution in database \cite{card}, as operations with lower cardinality can be prioritized to reduce intermediate result sizes and improve overall efficiency. For database (DB) APIs, cardinality can be estimated using metadata (\eg \text{min}, \text{max}, and \text{distinct\_count}) and the converted SQL query. Our system uses a simple estimation approach based on this metadata. For KG APIs and other APIs, if a \text{search\_key} is provided, the cardinality is estimated as \(1\), assuming the key uniquely identifies a single entity. If no \text{search\_key} is provided, the cardinality is estimated as \(\infty\), reflecting that we can't identify the entity based on the current interface.

Regardless of how the individual $GET$ and $JOIN$ are implemented, we propose a unified interface execution algorithm. The parsing algorithm can deal with $GET$s connected into chains, as well as the more general graph structure. Intuitively, we greedily select the $GET$ operator estimated to have the smallest cardinality to execute the whole interface more efficiently. The overall algorithm is presented in Algorithm~\ref{alg:interface_parsing}.

Initially, our algorithm initializes a priority queue and inserts all $GET$
nodes into the queue, prioritizing them based on their base cardinality. During each iteration, the $GET$ node $v_{min}$ with the smallest cardinality is extracted, and its $GET$ operation is executed on its own or with its recorded joining neighbor $v_{join}$, and the newly executed or combined node is added to the queue. We then identify join edges connected to the newly executed or combined node and perform join operations with its neighboring $GET$ nodes. After each join operation, the conditions and priorities of the neighbor nodes are updated. Specifically, we re-estimate the cardinality of joining the neighbor node and the combined node. If the combined cardinality is smaller than the value in the queue, we update the cardinality, and record which joined node $v_{join}$ the smaller cardinality is achieved. This process continues until only one result remains in the queue. By prioritizing nodes with smaller cardinality, the algorithm uses a greedy strategy to optimize execution.

\begin{algorithm}[t]
\caption{Efficient Execution of Chain of APIs.}
\label{alg:interface_parsing}
\LinesNumbered

\KwIn{API Chain \( Chain = (V, E) \), where \( V \) is the set of $GET$ nodes and  $E $ is the set of $JOIN$ edges.
}

\KwOut{Execution Result. }
 
Initialize priority queue \( Q \);

\ForEach{$ v \in V $}{
     $ c \gets \text{Estimate}(v) $;
     
     Insert $ v $ into $ Q $ with cardinality $ c $, initializing $ v[\text{join}] \gets \emptyset $;}

\While{$ \text{len}(Q) > 1 $}{
     Extract $ v_{\text{min}} $ with smallest cardinality from $ Q $;

     Create merged node $ v_{\text{new}} $ combining $ v_{\text{min}} $ and $ v_{\text{min}}[\text{join}] $ (if exists);
     
     Remove $ v_{\text{min}} $ and $ v_{\text{min}}[\text{join}] $ from $ Q $ if present;
     
     Insert $ v_{\text{new}} $ into $ Q $;
     
    \ForEach{edge $ e $ from $ v_\text{new} $ to neighbor $ v_{\text{n}} $}{
    
     Compute joined cardinality $ c_{\text{new}} \gets \text{EstimateJoin}(v_{\text{new}}, v_{\text{n}}, e) $;

     \If{$ c_{\text{new}} < $ current cardinality of $ v_{\text{n}} $}
     {
     Update $ v_{\text{n}} $'s cardinality in $ Q $;
     
     Record $ v_{\text{n}} $'s new $ v_{\text{n}}[\text{join}] \gets v_{\text{new}} $;
    }
}
}
\Return  The remaining node in $ Q $;
\vspace*{-0.1cm} 
\end{algorithm} 

\section{ER-RAG Pipeline}
\label{sec-errag}

In this section, we present the ER-RAG pipeline, a two-stage API generation module followed by a post-processing module. In the API generation module, the first stage focuses on source selection, where an agent fine-tuned using the DPO framework~\cite{dpo} evaluates and selects the most relevant sources based on task requirements. This stage aims to pass the data sources with sufficient information to answer the query and the least retrieval time to the second stage. In the second stage, a generation agent constructs the API chains by leveraging the schema of the selected sources as prompts, enabling precise and context-aware API generation. Following API retrieval, a post-processing module is employed to apply additional operations to the collected data, enhancing the expressive power and flexibility of the RAG system.

\subsection{API Generation Objective.}
 Given multi-source data  \( \mathcal{S} \) and an execution system P, we aim to construct an optimal API  chain  to achieve the dual objectives of \textbf{maximizing answer accuracy } and \textbf{minimum retrieval latency}. The multi-source data  \( \mathcal{S} \) can include several separate heterogeneous data sources, \eg DB, WIKI, WEB, etc, each containing different data and having different retrieval costs.
 \eat{
Given multi-source data and a parsing system, the goal is to generate an optimal API \( \sigma \) such that, supported by the parsing system and the multi-source data \( \mathcal{D} \), the LLM \( \mathcal{M} \) can produce the correct answer \( a^* \) with \textbf{maximum accuracy} and \textbf{minimum latency}.   The optimization objective can be expressed as:
\[\max_{\sigma} \left( P_{\mathcal{M}}(a^* \mid \sigma(q, \mathcal{D})) - \lambda \cdot T(\sigma(q, \mathcal{D})) \right),\]where \( \sigma(q, \mathcal{D}) \) represents the API generated for the target question \( q \), which leverages the support from the parsing system and multi-source data \( \mathcal{D} \). \( P_{\mathcal{M}}(a^* \mid \sigma(q, \mathcal{D})) \) is the probability of the LLM \( \mathcal{M} \) generating the correct answer \( a^* \) based on the API \( \sigma \). \( T(\sigma(q, \mathcal{D})) \) denotes the time cost of invoking the API, including the parsing system processing time, and multi-source data query time. Finally, \( \lambda \) is a hyperparameter to balance accuracy and time cost, where a higher \( \lambda \) prioritizes minimizing time cost, and a lower \( \lambda \) emphasizes accuracy. 
}

\subsection{API Generation for Provided Sources.} 

Since training the source selection module requires generating and evaluating APIs on fixed sources, we will first present API generation for given sources. Given a data source and a query, we propose combining prompt engineering and tuning to generate the API. The prompt is as follows:
\vspace{-0.2cm}
\begin{lstlisting}[label={lst:api_generation}]
You are an agent for building a retrieval system.
### Instruction: 
You need to generate an API with *{Datasource_name}*.
### API Definition: @{API_definition}@  
### Datasource Schema: ^{Datasource_schema}^ 
### Query:
Please refer to the instructions and  generate an API for the following query directly:
Query: +{Query}+
Interface:
\end{lstlisting}
The prompt is unified for all sources in the \{API\_definition\} part, which describes the API format. The \{Datasource\_name\} is the name of the provided source, including single sources, \eg SELF, DB\_Finance, WIKI, WEB, etc, and mixed sources, \eg WIKI +  DB\_Fi-nance, etc. For different data sources, the \{Datasource\_schema\} part is different. For example, for the database part, we have to fill in the \{Datasource\_schema\} part with the related database schema, otherwise, it would be impossible for the LLM to generate the correct table name or row name.  For data sources such as the WEB  or Wikipedia, since the LLM already has knowledge of these sources, we only need to provide it with the names of the data sources.  For mixed sources, the \verb|{Datasource_schema}| part should include all relevant component schemas.  

Given fixed data sources, we prompt GPT-4o to generate the reference APIs from the training set and then correct them manually as the ground-truth data. We then fine-tune a small LLM (8B) using the annotated interfaces to perform the interface generation task.

\subsection{Source Selection}

Selecting different data sources is essential for ensuring accurate and efficient responses. Firstly, similar to the schema linking module in NL2SQL, we need to choose the right source in homogeneous sources. For example, if we have two DB sources, DB\_Music and DB\_Movie, and we are provided with a question related to music, we need to direct the source correctly to music, otherwise, the other data source contains no useful information. Moreover, heterogeneous sources have their strengths and weaknesses. If a wrong source is selected or all sources are simply stacked together without consideration, it can lead to incorrect or suboptimal answers. Proper selection ensures that the most suitable source combination is chosen for a given task, leveraging its strengths while avoiding unnecessary conflicts.

Secondly, different sources or models have varying execution times. For example, only considering the time cost, using a search engine to search and fetch the top 10 related web pages may cost 2 seconds, and searching from a database may cost milliseconds to hundreds of seconds or more depending on the database scale and query difficulty, and searching from Wikipedia and parsing the entity and edges using LLMs may cost a few seconds to hundreds of seconds depending on how deeply the LLM is involved. To provide timely responses, it's crucial to assign queries to the source with the shortest execution time that can still deliver accurate results. 

In this subsection, we present our source selection training pipeline, consisting of a supervised fine-tuning (SFT) training part that teaches the LLM which sources can accurately capture the related information, and a direct policy optimization (DPO) preference training part that guides the LLM to choose the source with the lowest cost. Our experiments show that directly using SFT to select the single correct and fastest source leads to significant issues. Specifically, SFT struggles to balance multiple objectives effectively, causing the model to confuse whether a source is valid or simply low-cost. This confusion results in an over-selection of the SELF source (the source not using RAG with 0 retrieval cost), which ultimately harms performance. To address this, we chose to use DPO for further fine-tuning of the LLM. DPO avoids the pitfalls of SFT by better accommodating multiple objectives.
\begin{figure} [t]
    \centering
    \includegraphics[width=\columnwidth]{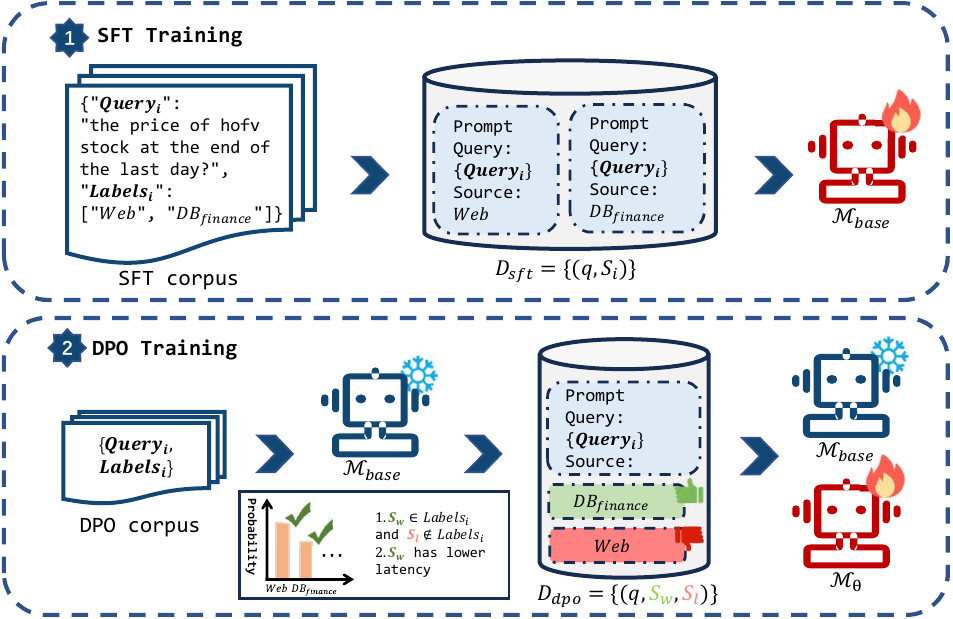}
    \vspace{-0.5cm}
    \caption{Fine-Tuning Source Selection Module Illustration.}
    \label{fig:task3framework}
    \vspace{-0.5cm}
\end{figure}

\stitle{SFT Training.} In order to help the LLM learn which source is reliable, we introduce SFT training before DPO. This stage helps the model output in categorical format and spot the reliable data source. The prompt we use is as follows:\noindent
\vspace{-0.2cm}
\begin{lstlisting}[label={lst:routing_agent}]
You are an agent for selecting the appropriate data source to answer user queries.
### Instruction:
Based on the provided query and available data sources (*{Datasource_list}*), you must determine the best option for obtaining the information needed.
### Source Schema Overview:^{Source_description}^
### Query:
Please refer to the instructions and output a most likely data source for the following query directly.
Query: +{Query}+
Source:
\end{lstlisting}
where $\{\text{Datasource\_list}\}$ is a list of sources, \eg DB\_movie, DB\_music, WIKI, WEB, SELF, MIX, etc. SELF means that no information retrieval is needed. MIX means allowing the API to query multiple sources. Each mix source is assigned to a special token in source selection tuning. $\{\text{Source}\_\text{description}\}$ contains a brief schema and source description of each source. For example, for a specific DB source, the description contains lists of table names and usages.

Here, given a query $q$, we denote the probability of LLM $\mathcal{M}$ choosing source $S$ as $P_{\mathcal{M}}(S|q)$. We construct the training set as follows: For each query $q$ in the training set $\mathcal{T}$, we generate an API using source $q$. Then, we execute the API and infer using the LLM with the retrieved context. Finally, we construct a set $\mathcal{D}_{sft}=\{(q,S_i)\}$, where query $q$ is answered correctly with the help of source $S_i$. Here, each $q$ can form pairs with multiple correct sources. We optimize the following SFT loss on $\mathcal{D}_{sft}$:
\vspace{-0.1cm}
\begin{equation*}
\mathcal{L}_{\text{SFT}} = 
- \mathbb{E}_{(q,S_i) \sim \mathcal{D}_{sft}} 
\left[
P_{\mathcal{M}}(S_i|q)
\right].   
\vspace{-0.1cm}
\end{equation*}
Training on this loss helps the LLM learn which external sources are helpful in the RAG or whether the internal knowledge inside the LLM is sufficient to answer the query.

\stitle{DPO Training.} We leverage DPO training to directly optimize the source selection agent according to the preference data. We adopt a strategy to rank the sources based on execution time. Such a strategy aims to train the model to choose the source with a lower running time, which can boost the efficiency of the retrieval process.

Motivated by the online preference data construction~\cite{dpo_kim2025spread}, we generate the preference training set $\mathcal{D}_{dpo}$ as follows: After the SFT training step, we obtain a base reference model $\mathcal{M}_{base}$. For each query $q$ in the training set $\mathcal{T}$, we use $\mathcal{M}_{base}$ to infer the source $S_1$ and $S_2$ with the top-2 probabilities. For the execution time strategy, based on the source retrieval time, if source $S_1$ is correct and has the lowest retrieval time, we add the winning source $S_w=S_1$ and lossing source $S_l=S_2$ into the preference training set $\mathcal{D}_{dpo}$. Otherwise, we identify the correct source with the lowest retrieval time $S_{min}$, and we add $S_w=S_{min}>S_l=S_1$ into  $\mathcal{D}_{dpo}$. This strategy allows us to update the current reference model marginally, helping the model to capture the retrieval time pattern to the largest extent.
We use the following DPO loss~\cite{dpo} to get the final module $\mathcal{M}_\theta$:
\vspace{-0.1cm}
\begin{equation*}
\begin{aligned}
&\mathcal{L}_{\text{DPO}}(\mathcal{M}_\theta; {\mathcal{M}_{base}}) = 
- \mathbb{E}_{(q, S_w, S_l) \sim \mathcal{D}_{\text{dpo}}}  \\& 
\bigg[
\log \sigma \bigg( 
\beta \log \frac{P_{\mathcal{M}_{\theta}}(S_w \mid q)}{P_{\mathcal{M}_{base}}(S_w \mid q)} -  \beta \log \frac{P_{\mathcal{M}_{\theta}}(S_l \mid q)}{P_{\mathcal{M}_{base}}(S_l \mid q)} 
\bigg)
\bigg],
\vspace{-0.1cm}
\end{aligned}
\end{equation*}
where the current model $\mathcal{M}_\theta$ is initialized by the base model $\mathcal{M}_{base}$, $\beta$ is a parameter controlling the deviation from the base reference model, and $\sigma$ is the sigmoid function. 

\subsection{Post-Processing}

Our unified API efficiently extracts relevant information from multiple data sources, but certain queries—such as those involving superlatives, ordinal rankings, or numerical computations (\eg "What’s the latest film directed by Walt Becker?")—require post-processing to deliver accurate results. While basic operations like $GET$ and $JOIN$ handle data retrieval and integration, they lack the ability to perform advanced tasks like sorting, grouping, or numerical computations, which are essential for operations akin to ORDER BY or GROUP BY in databases. Post-processing addresses these limitations by enabling transformations, comparisons, and aggregations on the result set after initial API data retrieval.

Additionally, post-processing can leverage LLMs to help data normalization, relevance verification, and string modification.  For data normalization, post-processing standardizes formats to enable seamless comparison and sorting. For example, it unifies time formats (\eg converting "2023-10-01" and "October 1, 2023" into a single standard) and unit measurements (\eg converting "5 km" to "5000 meters").  For relevance verification, post-processing leverages LLMs to assess the alignment between retrieved content and the user’s query. By analyzing semantic cues, contextual information, and user intent, it filters out irrelevant data and ensures that only pertinent information is considered for further processing.  For string manipulation, post-processing refines raw outputs by transforming attribute-value pairs into clear, natural language formats. These capabilities ensure that our system delivers accurate, interpretable, and user-friendly results across a wide range of queries.

To enable flexible and efficient post-processing, we propose generating lightweight Python f-string scripts. An f-string is a concise and powerful way to embed expressions directly into string literals. The f-string takes the list of dictionary outputs of the API retrieval result and uses the braced expressions to process the result. Unlike traditional string formatting methods, f-strings allow for dynamic evaluation of embedded code and they can handle heterogeneous formats like units or time representations well. For example, for a query like \textit{"What are the top 3 highest revenues among these companies?"}, the retrieved data may include values with mixed units (\eg [\{company:"Company A",revenue:"120M"\}, \{company:"Company B",revenue:"8500K"\},...]). A one-line f-string script normalizes these values into a consistent unit (\eg millions) and computes the result:
\vspace{-0.1cm}
\begin{lstlisting}
f%'''The top 3 companies by revenue are% !{", ".join([x["company"] for x in sorted(Data, key=lambda x: float(x["revenue"].replace("M", "")) if "M" in x["revenue"] else float(x["revenue"].replace("K", ""))/1000, reverse=True)[:3]])}!%.'''%
\end{lstlisting}
\noindent This approach sorts the data and extracts the top three entries directly, ensuring precise results and avoiding LLM hallucination.

We prompt and tune the LLM to generate the concise Python script for post-processing. The generation prompt is as follows:
\vspace{-0.1cm}
\begin{lstlisting}[label={lst:structured_data_processing}]
You are an agent for processing structured data.
### Instructions:
Extract useful information from the retrieved data that directly addresses the user's query.  
If the task involves computation or dynamic formatting, use Python's "f-string" to embed expressions or calculations.   
### API Definition: @{API_definition}@ 
### Output Requirements:
Complete the Return part in the following Python function template:  
!def solve(Data):
    from datetime import datetime
    return {{Return}}!
Replace {{Return}} with your generated f-string or "no data" if no relevant data is found.
### Query:
Question: +{query}+
API: +{api}+
Data (Limited by maximum length): +{data}+
Return: f'''
\end{lstlisting}
Here, \{API\_definition\} is a brief description of the API rules helping the LLM to understand the API format. Using this prompt, we generate the Python scripts for the samples in the training set using GPT-4o, and we tune an agent for the post-processing task.

\section{Experiment}
\label{sec-exp}

In this section, we conduct extensive experiments to evaluate ER-RAG. Our objective is to address the following research inquiries through our experiments:

\hangafter 1
\hangindent 2em
I1: How effectively does our interface handle individual sources, and how does it integrate information across diverse sources?

\hangafter 1
\hangindent 2em
I2: How does our API enhance multi-source multi-hop queries and support agent construction on new data sources?

\hangafter 1
\hangindent 2em
I3: How does ER-RAG generalize out of the training set? 

\hangafter 1
\hangindent 2em
I4: How does source selection affect the performance and efficiency of RAG on multiple sources? 
\eat{
\begin{itemize}
    \item I1: How does our interface work on individual sources? How does our unified interface help integrate information from different sources?     
    \item I2: How does our API benefit multi-source multi-hop queries? How does the API design benefit agent constructions on new data sources?
    \item I3: How does each module generalize out of the training set? 
    \item I4: How does source selection affect the performance and efficiency of RAG on multiple sources? 
\end{itemize}

Regarding I1, we show the comparison of ER-RAG with other single-source RAG and hybrid-source RAG baselines on the CRAG dataset in Sec. 5.2. Regarding I2, we present ER-RAG's capability on multi-source multi-hop queries on the synthetic MIX dataset in Sec.5.3, and we study the effect of joint training of API generation to illustrate the benefit of unified agent construction samples of one source on another. Regarding I3, we show our performance on the Compmix dataset with ER-RAG trained on the CRAG training set to show the generalization of ER-RAG in Sec. 5.4. Regarding I4, in Sec. 5.5, we show the performance and time cost of different source selection methods.
}

To address \textbf{I1}, we compare ER-RAG with single-source and hybrid-source RAG baselines on the CRAG dataset in Sec. 5.2. For \textbf{I2}, we evaluate ER-RAG's performance on multi-source multi-hop queries using the synthetic MIX dataset in Sec. 5.3, and analyze the impact of joint API generation training to demonstrate the benefits of unified agent construction across sources. Regarding \textbf{I3}, we assess ER-RAG's generalization capability by testing its performance on the Compmix dataset (trained on the CRAG training set) in Sec. 5.4. Finally, for \textbf{I4}, we present the performance and time cost of various source selection methods in Sec. 5.5.

\begin{table}[t]
    \centering
    \label{tab:dataset}
    \caption{Statistics of the Datasets.}
    \vspace{-0.3cm}
    \begin{tabular}{l|c|c}
        \toprule
        \textbf{Train} & \textbf{\#Samples} & \textbf{Avg. Tokens} \\
        \midrule
        CRAG-API-Train& 3,901& 1,116\\
        CRAG-Post-Train& 2,082& 636\\
        CRAG-Entity-Train &4,921  &1,851 \\
        Synthetic-API-Train & 726& 2,519 \\
        \midrule
        \textbf{Test} & \textbf{\#Questions
} & \textbf{Avg. Tokens} \\
        CRAG-Test& 1,212& 15.6\\
        CompMix  & 1,531& 11.4\\
        Synthetic-Test& 916 & 43.9\\
        \midrule
        \textbf{Database} & \textbf{\#Tables} & \textbf{Avg. Columns/Rows} \\
        Movie Database &7 & 7/243,313\\
        Music Database& 5& 7/1,116,301\\
        Sports Database& 2& 44/33,915\\
        Finance Database&6 &28/1,142,958 \\
        Open Database& 1& 5/3,355\\
        \midrule
        \textbf{Other Sources} & \textbf{\#Items} & \textbf{Avg. Tokens} \\
        Cached WEB Page &68,079 & 9,021\\
        Cached WIKI Page & 4,740&6,436 \\
        \bottomrule
    \end{tabular}
    \label{tab:dataset}
\vspace{-0.4cm}
\end{table}

\definecolor{mygreen}{RGB}{34,145,34}
  \begin{table*}[]
\centering
\caption{RAG Performance on the CRAG Dataset (Colored numbers indicate the best results among a certain type of methods).}
\vspace{-0.3cm}
\label{tab:crag}
\resizebox{\textwidth}{!}{%
\begin{tabular}{c|l|llllll|llllll}
\toprule
\multirow{2}{*}{Types}& \multicolumn{1}{c|}{\multirow{2}{*}{Methods}} & \multicolumn{6}{c|}{LLM scores} & \multicolumn{6}{c}{Accuracy}\\ \cline{3-14} 
& \multicolumn{1}{c|}{} & Movie & Music & Finance & Sports& Open& All & Movie & Music & Finance & Sports& Open& All \\ \midrule
\multirow{4}{*}{\begin{tabular}[c]{@{}c@{}}LLM \\ only\end{tabular}}& Llama3 8B &{35.1}\ ($\downarrow$33.0)& 38.5\ ($\downarrow$20.7)& 14.5\ ($\downarrow$32.2)& 25.0\ ($\downarrow$28.7)& 51.1\ ($\downarrow$14.4)& 31.4\ ($\downarrow$26.9)& 14.7\ ($\downarrow$31.9)& 23.0\ ($\downarrow$6.3)& 1.9\ ($\downarrow$31.5) & {10.6\ ($\downarrow$27.4)}& 10.0\ ($\downarrow$30.7)& 10.9\ ($\downarrow$27.1)\\
& GPT-4o-mini& 45.5\ ($\downarrow$22.6)&52.3\ ($\downarrow$6.9)&14.2\ ($\downarrow$32.5)& 40.7\ ($\downarrow$13.0)&60.7\ ($\downarrow$4.8)&40.3\ ($\downarrow$18.3)&\makecell{26.2\ ($\downarrow$20.4)}&21.8\ ($\downarrow$7.5)& 6.6\ ($\downarrow$26.8) & 21.3\ ($\downarrow$16.7)& 33.2\ ($\downarrow$7.5)&20.9\ ($\downarrow$17.1)\\
& GPT-4o &\textcolor{blue}{\textbf{57.7}}\ ($\downarrow$10.4) &53.4\ ($\downarrow$5.8)&13.2\ ($\downarrow$33.5)& 50.9\ ($\downarrow$2.8)& 66.8\ ($\uparrow$1.3)& 46.0\ ($\downarrow$12.3)& 30.8\ ($\downarrow$15.8)& 25.9\ ($\downarrow$3.4)& 6.6\ ($\downarrow$26.8)& 26.4\ ($\downarrow$11.6)& \textbf{38.4}\ ($\downarrow$2.3) & 24.4\ ($\downarrow$13.6)\\
& DeepSeek v3& 53.4\ ($\downarrow$14.7)& \textcolor{blue}{\textbf{55.2}\ ($\downarrow$4.0)} & \textcolor{blue}{\textbf{18.6}\ ($\downarrow$28.1)} & \textcolor{blue}{\textbf{59.7}}\ ($\uparrow$6.0) & \textcolor{blue}{\textbf{67.7}}\ ($\uparrow$2.2) & \textcolor{blue}{\textbf{48.4}\ ($\downarrow$9.9)} & \textcolor{blue}{\textbf{33.3}\ ($\downarrow$13.3)} & \textcolor{blue}{\textbf{27.6}\ ($\downarrow$1.7)} & \textcolor{blue}{\textbf{12.0}\ ($\downarrow$21.4)} & \textcolor{blue}{\textbf{31.9}\ ($\downarrow$6.1)} & \textcolor{blue}{\textbf{38.4}\ ($\downarrow$2.3)} & \textcolor{blue}{\textbf{27.7}\ ($\downarrow$10.3)} \\ \midrule
\multirow{3}{*}{\begin{tabular}[c]{@{}c@{}}Industry \\ systems\end{tabular}}& ChatGPT Plus (GPT-4o)& 57.7\ ($\downarrow$10.4)& 56.3\ ($\downarrow$2.9)& \textcolor{violet}{\textbf{22.4}\ ($\downarrow$24.3)} & 50.9\ ($\downarrow$2.8)& 68.6\ ($\uparrow$3.1)& 49.1\ ($\downarrow$9.2)& \textcolor{violet}{\textbf{34.8}\ ($\downarrow$11.8)} & 23.6\ ($\downarrow$5.7)& \textcolor{violet}{\textbf{14.8}\ ($\downarrow$18.6)} & \textcolor{violet}{\textbf{29.2}\ ($\downarrow$8.8)} & \textcolor{violet}{\textbf{40.2}\ ($\downarrow$0.5)} & \textcolor{violet}{\textbf{28.0}\ ($\downarrow$10.0)} \\
& Claude-3-5-sonnet Pro & 53.0\ ($\downarrow$15.1)& 59.2\ ($\downarrow$0.0)& 15.5\ ($\downarrow$31.2)& 52.3\ ($\downarrow$1.4)& 70.3\ ($\uparrow$4.8)& 47.2\ ($\downarrow$11.1)& 29.7\ ($\downarrow$16.9)& \textcolor{violet}{\textbf{27.0}\ ($\downarrow$2.3)} & 9.5\ ($\downarrow$23.9) & 27.3\ ($\downarrow$10.7)& 38.9\ ($\downarrow$1.8)& 25.3\ ($\downarrow$12.7)\\
& Perplexity (405B) & \textcolor{violet}{\textbf{68.5}\ ($\uparrow$0.4)} &\textcolor{violet}{\textbf{62.1}\ ($\uparrow$2.9)} & 18.6\ ($\downarrow$28.1)& \textcolor{violet}{\textbf{60.2}\ ($\uparrow$6.5)} & \textcolor{violet}{\textbf{80.8}\ ($\uparrow$15.3)} & \textcolor{violet}{\textbf{55.4}\ ($\downarrow$2.9)} & 30.1\ ($\downarrow$16.5)& 26.4\ ($\downarrow$2.9)& 11.7\ ($\downarrow$21.7)& 24.5\ ($\downarrow$13.5)& 31.0\ ($\downarrow$9.7)& 24.0\ ($\downarrow$14.0)\\ \midrule
\multirow{7}{*}{\begin{tabular}[c]{@{}c@{}}Single \\ data source \\ (Llama3 8B)\end{tabular}} & ZeroShot SQL (DB,GPT-4o) & 54.8\ ($\downarrow$13.3)& 36.8\ ($\downarrow$22.4)& 24.9\ ($\downarrow$21.8)& 12.5\ ($\downarrow$41.2)& 21.4\ ($\downarrow$44.1)& 30.6\ ($\downarrow$27.7)& 35.8\ ($\downarrow$10.8)& 16.7\ ($\downarrow$12.6)& 18.3\ ($\downarrow$15.1)& 8.8\ ($\downarrow$29.2) & 14.0\ ($\downarrow$26.7)& 19.6\ ($\downarrow$18.4)\\
& DAIL SQL (DB,GPT-4o) & 62.0\ ($\downarrow$6.1)& 43.1\ ($\downarrow$16.1)& 35.6\ ($\downarrow$11.1)& 38.9\ ($\downarrow$14.8)& 38.9\ ($\downarrow$26.6)& 44.0\ ($\downarrow$14.3)& 39.4\ ($\downarrow$7.2)& 19.5\ ($\downarrow$9.8)& 24.9\ ($\downarrow$8.5)& 28.7\ ($\downarrow$9.3)& 23.6\ ($\downarrow$17.1)& 27.9\ ($\downarrow$10.1)\\
& SuperSQL (DB,GPT-4o) & \textcolor{orange}{\textbf{63.4}\ ($\downarrow$4.7)} & 42.5\ ($\downarrow$16.7)& 37.5\ ($\downarrow$9.2)& \textcolor{orange}{\textbf{49.1}\ ($\downarrow$4.6)} & 40.6\ ($\downarrow$24.9)& 46.8\ ($\downarrow$11.5)& 43.0\ ($\downarrow$3.6)& 20.7\ ($\downarrow$8.6)& 26.5\ ($\downarrow$6.9)& \textcolor{orange}{\textbf{34.7}\ ($\downarrow$3.3)} & 23.6\ ($\downarrow$17.1)& 30.4\ ($\downarrow$7.6)\\
& ER-RAG (DB) & 61.7\ ($\downarrow$6.4)& 47.1\ ($\downarrow$12.1)& \textcolor{orange}{\textbf{44.8}\ ($\downarrow$1.9)} & 40.3\ ($\downarrow$13.4)& 47.6\ ($\downarrow$17.9)& \textcolor{orange}{\textbf{48.7}\ ($\downarrow$9.6)} & \textcolor{orange}{\textbf{47.0}\ ($\uparrow$0.4)} & 24.7\ ($\downarrow$4.6)& \textcolor{orange}{\textbf{32.1}\ ($\downarrow$1.3)} & 31.9\ ($\downarrow$6.1)& 30.3\ ($\downarrow$10.4)& \textcolor{orange}{\textbf{34.1}\ ($\downarrow$3.9)} \\
& ER-RAG (WEB Chunk)& 53.0\ ($\downarrow$15.1)& 50.6\ ($\downarrow$8.6)& 19.2\ ($\downarrow$27.5)& 42.6\ ($\downarrow$11.1)& \textcolor{orange}{\textbf{65.9}\ ($\uparrow$0.4)} & 44.4\ ($\downarrow$13.9)& 35.1\ ($\downarrow$11.5)& \textcolor{orange}{\textbf{27.0}\ ($\downarrow$2.3)} & 13.2\ ($\downarrow$20.2)& 24.5\ ($\downarrow$13.5)& \textcolor{orange}{\textbf{38.9}\ ($\downarrow$1.8)} & 27.1\ ($\downarrow$10.9)\\
& ER-RAG (WEB ER) & 51.3\ ($\downarrow$16.8)& \textcolor{orange}{\textbf{53.4}\ ($\downarrow$5.8)} & 21.1\ ($\downarrow$25.6)& 40.3\ ($\downarrow$13.4)& 59.4\ ($\downarrow$6.1)& 43.3\ ($\downarrow$15.0)& 35.1\ ($\downarrow$11.5)& 20.1\ ($\downarrow$9.2)& 14.0\ ($\downarrow$19.4)& 21.8\ ($\downarrow$16.2)& 32.9\ ($\downarrow$7.8)& 24.7\ ($\downarrow$13.3)\\
& ER-RAG (WIKI ER)& 49.5\ ($\downarrow$18.6)& 50.6\ ($\downarrow$8.6)& 11.4\ ($\downarrow$35.3)& 28.7\ ($\downarrow$25.0)& 62.9\ ($\downarrow$2.6)& 38.5\ ($\downarrow$19.8)& 33.0\ ($\downarrow$13.6)& 21.8\ ($\downarrow$7.5)& 7.9\ ($\downarrow$25.5) & 15.7\ ($\downarrow$22.3)& 35.4\ ($\downarrow$5.3)& 22.2\ ($\downarrow$15.8)\\ \midrule
\multirow{4}{*}{\begin{tabular}[c]{@{}c@{}}Hybrid \\ data sources\\ (Llama3 8B)\end{tabular}}  
& ER-RAG (MIX)& 61.7\ ($\downarrow$6.4)& 48.3\ ($\downarrow$10.9)& 43.5\ ($\downarrow$3.2)& 45.4\ ($\downarrow$8.3)& 56.8\ ($\downarrow$8.7)& 51.2\ ($\downarrow$7.1)& 45.5\ ($\downarrow$1.1)& 26.4\ ($\downarrow$2.9)& 20.6\ ($\downarrow$12.8)& 31.0\ ($\downarrow$7.0)& 30.3\ ($\downarrow$10.4)& 30.8\ ($\downarrow$7.2)\\
& ER-RAG (All in Context)& 65.9\ ($\downarrow$2.2)& 54.6\ ($\downarrow$4.6)& 41.0\ ($\downarrow$5.7)& \textcolor{mygreen!110}{\textbf{57.4}\ ($\uparrow$3.7)} & 60.3\ ($\downarrow$5.2)& 55.2\ ($\downarrow$3.1)& 46.2\ ($\downarrow$0.4)& 24.7\ ($\downarrow$4.6)& 27.9\ ($\downarrow$5.5)& 36.1\ ($\downarrow$1.9)& 35.1\ ($\downarrow$5.6)& 34.5\ ($\downarrow$3.5)\\
& ER-RAG (Source Selection)& \textcolor{mygreen!110}{\textbf{68.1}} & \textcolor{mygreen!110}{\textbf{59.2}} & \textcolor{mygreen!110}{\textbf{46.7}} & 53.7& \textcolor{mygreen!110}{\textbf{65.5}} & \textcolor{mygreen!110}{\textbf{58.3}} &\textcolor{mygreen!110}{ \textbf{46.6}} &\textcolor{mygreen!110}{\textbf{29.3}} &\textcolor{mygreen!110}{\textbf{33.4}} & \textcolor{mygreen!110}{\textbf{38.0}} & \textcolor{mygreen!110}{\textbf{40.7}} & \textcolor{mygreen!110}{\textbf{38.0}} \\
& \textcolor{gray}{ER-RAG (Oracle)} &  \textcolor{gray}{81.7}& \textcolor{gray}{76.6}&  \textcolor{gray}{58.4}& \textcolor{gray}{72.7}& \textcolor{gray}{82.5}& \textcolor{gray}{73.4}& \textcolor{gray}{53.8}& \textcolor{gray}{35.6}&\textcolor{gray}{36.2}& \textcolor{gray}{43.9}& \textcolor{gray}{45.5}& \textcolor{gray}{43.3}\\ \bottomrule
\end{tabular}%
}
\vspace{-0.3cm}
\end{table*}

\subsection{Experimental Setup}
\stitle{Datasets.} To study our RAG system on heterogeneous data, we conduct experiments on CRAG, CompMix, and synthetic MIX datasets.

The CRAG dataset~\cite{yang2024crag} is a comprehensive benchmark for multi-source RAG systems, consisting of QA pairs spanning multiple domains and diverse question types. CRAG simulates real-world retrieval scenarios by providing retrieval content from web searches and knowledge graphs (KG), including up to 50 HTML web pages and a KG equipped with Mock API with 2.6 million entities. In the original dataset, the Mock API allows querying entities in a black-box manner, which is only accessible by entity names. We implement three API sources for CRAG: DB, WIKI, and WEB. For the DB source, we re-organize and unify the Mock API data into a PostgreSQL database, turning it into a "white-box" source for easier comparison. The WIKI source relies on an online wiki parsing system we implement, while the WEB source uses the original 50 web pages.   Since the test set of CRAG is not open-sourced, we  use split=0 from its pre-divided splits in the open-source dataset as the test set and split=1 as the training set. 

CompMix~\cite{CompMix} evaluates QA methods over heterogeneous sources like WIKIDATA, Wikipedia text, tables, and infoboxes. We also implement the same three API sources for CompMix. For the DB source, we use the music, movie, and soccer CRAG DB. We use the same WIKI parsing system and implement a web-fetching system for WEB sources using Google Search APIs, retrieving 5 pages per query. Testing on CompMix demonstrates our system's robustness and generalization to new questions and sources, with modules tuned only on CRAG.

To demonstrate our method's performance on multi-source retrieval, we construct a synthetic multi-source query set. For mixed sources, to create a training set of source $S_1$ and $S_2$, we follow these steps: 1. Select queries from the CRAG dataset answerable using $S_1$ as step-one questions, filtering for entity answers so the first query's answer can serve as part of the join condition for the second query. 2. Prompt LLM to generate similar queries for $S_2$ using correctly answered examples from $S_2$. Execute the same retrieval pipeline to extract context from $S_2$ and prompt LLM to check context relevance and generate a second-step query. 3. For related second-step queries, prompt LLM to generate a combined two-step query with the first-step query, excluding the join condition entity. We prompt LLM to generate a unified API to form the mixed-source training set. We construct the synthetic dataset on WIKI and DB mixed sources. We split the synthetic dataset into training and test sets based on the source of the step-one question, ensuring that questions from the CRAG training set are in the training set and those from the CRAG test set are in the test set.

We present the statistics of these three datasets in Table~\ref{tab:dataset}. Specifically, in the training set, CRAG-API-train refers to a training set of samples generated by a single source API, while Synthetic-API-Train is a mixed multi-source training sample. CRAG-Post-Train refers to the training set for the post-processing Python script, and CRAG-Entity-Train refers to the training set for the module responsible for entity disambiguation and relation extraction in our WIKI/WEB source code implementation.

\stitle{Metrics.} To evaluate QA accuracy, two common metrics can be used: \textbf{Stem-based Accuracy}~\cite{selfrag} (Accuracy in Table~\ref{tab:crag}) and \textbf{LLM-based Scores}~\cite{chen2019evaluating,yang2024crag}. Stem-based accuracy focuses on stemmed tokens. If all stemmed tokens from the ground truth are present in the predicted output, the answer is labeled correct. Otherwise, the answer is considered wrong.

For LLM-based Score, an LLM is used to determine whether the predicted answer is semantically correct compared to the ground truth. The model outputs a True (if the answer is correct) or False (if the answer is incorrect), providing a binary evaluation for accuracy.  We use the prompt from the CRAG contest~\cite{yang2024crag} to evaluate answer correctness using DeepSeek V3.

To evaluate the performance of the source selection system, we use accuracy and time cost for the information retrieval system to evaluate the effectiveness and efficiency of the RAG system. We use the \textbf{LLM score} as the accuracy metric, and we use the end-2-end \textbf{information retrieval time} as the time cost metric.

\eat{To evaluate the performance of the routing system, we need to count the AC (actually correct), the correct cases using our routing system, NR (correct (no RAG)), the model's capability of answering without using a rag system, BC (theoretical best correct), the combined correct cases from different sources. From the time perspective, we report the actual cost of our retrieval system and the theoretical maximum cost of our retrieval system, as we set a maximum search time for each query. To show the balance of efficiency and effectiveness, we use a mixed metric of 
\[
\text{Metric} = \frac{\text{AC} - \text{NR}}{\text{BC} - \text{NR}} \times \left(1 - \frac{\text{Actual Cost}}{\text{Theoretical Max Cost}}\right)
\]
where the first term is a normalized metric of how close the system's accuracy is to the theoretical best possible outcome, and the second term is also a normalized metric that measures the efficiency of the system in terms of cost. For No RAG system, the mixed metric is equal to 0. The mixed metric is closer to 1 for more ideal routing systems.}

\stitle{Baselines.}  For the retrieval performance, we compare the baseline methods from three different aspects. First, we test the basic capabilities of LLM without RAG on these questions to illustrate the effect of RAG on this task.

Second, we compare the performance of our whole system with industry systems on the same queries. We select ChatGPT Plus, Claude-3.5-sonnet Pro, Perplexity ai-Llama3.1 (405B), and three commercial RAG systems based on strong LLMs, as our baselines. The industry system can have access to various information sources and may be constructed on more powerful agents or models. 

Third, we compare our method with other baselines on single data sources. For instance, on the DB source, we select DAIL (VLDB'24)~\cite{gao2024text}, SuperSQL (VLDB'24)~\cite{NLP2SQLDawn} as the NL2SQL baseline, two LLM-based NL2SQL methods, which are another major approach to  generate  SQL queries to extract information from data sources. We implement them using GPT-4o as the SQL generation agent. \eat{For the web page source, we use the traditional content retrieval pipeline as a baseline (WEB Chunk).}

Finally, we evaluate our integration strategy on hybrid data sources by comparing ER-RAG (MIX) for mixed data, ER-RAG (Source Selection) with a source selection pipeline, and ER-RAG (All in Context) using all retrieved information. We also demonstrate the ideal performance of ER-RAG (Oracle), which assumes perfect source selection at minimal cost.

\stitle{Implementation Details.} We use Llama3 8B as our base model in all tests and also implement all the modules using it. The calls to the remaining LLMs are implemented through the OpenAI library of a third-party agent, with the temperature set to 0. We use the TRL library and PEFT library~\cite{wolf2019huggingface} to conduct LoRA fine-tuning on the LLMs. Specifically, we train four independent modules (API Generation, WIKI/WEB Entity Processing, Post-processing, and Source Selection), each corresponding to a separate LoRA adapter. During inference, these adapters can be dynamically switched  based on the task requirements. The LoRA configs are $lora\_alpha = 16, lora\_dropout = 0.1$, and $lora\_rank = 8$. 

For supervised fine-tuning, we train our model using the AdamW optimizer~\cite{loshchilov2017decoupled}. The number of training epochs is set to $2$ ($5$ for API generation), the learning rate is set to $2e-4$, and the batch size is set to $4$. For direct policy optimization, we employ the paged AdamW optimizer with $32$-bit precision, configured with a training duration of $1$ epoch and a learning rate of $1e-5$. The $\beta$ hyperparameter is tuned via a grid search over the values $\{0.8, 1, 1.2\}$ to identify the optimal setting for the task. In our ablation analysis, we replace fine-tuning with in-context learning~\cite{icl_dong2024survey} by preparing a prompt that includes the 5 problem-specific examples in the training set with the highest relevance scores.

\stitle{Relationship with the KDDCup Contest.} In the 2024 KDDCup Contest, we were provided with a MockAPI as the information source, which, given an entity name, returned nested JSON responses. Due to the limited number of LLM calls allowed in the competition and the relatively fixed output format of the MockAPI, we manually crafted rules to convert the MockAPI into the $GET$ API proposed in this paper~\cite{xia2024}.  In this work, we propose the $GET$ and $JOIN$ API chains, use the APIs to access all data sources (including newly expanded WIKI data source), refine API generation, and introduce source selection. The contest solution is less general and expressive than the ER-RAG framework proposed in our work.

\subsection{Overall Performance Comparison}

In this subsection, we compare ER-RAG with the baselines on the CRAG dataset to illustrate ER-RAG's performance on both single and hybrid sources. We present the main result of the CRAG dataset in Table~\ref{tab:crag}. The conclusions can be drawn from the results:

1. RAG shows strong performance on the CRAG dataset, achieving improvements of  \textbf{17.3\%}, \textbf{7.1\%}, and \textbf{13.0\%} in LLM scores over the Llama3 8B baseline for DB, WIKI, and WEB, respectively.

2. The API implemented by our RAG method performs comparably to other approaches, such as the NL2SQL or traditional web page retrieval method, on single data sources. Notably, the individual GET implementation, which is not our main proposal, can take advantage of the advances in each subtask.

3. The mixed-source agent (ER-RAG (MIX)) trained on a synthetic training set outperforms all the single-source agents, illustrating that using different sources in different hops in multi-hop questions is useful in the CRAG dataset.

4. Aggregating all sources together (ER-RAG (Oracle)) can unleash a total potential of \textbf{73.4\%} in LLM score, demonstrating the significant potential of the approach. However, fitting all the retrieval information from different sources in the LLM context (ER-RAG (All in Context)) results  in a performance loss of \textbf{22.2\%} compared to the whole potential, demonstrating that simply putting all the sources in the context can result in performance degradation. The complete ER-RAG with the  source selection module (ER-RAG (Source Selection)) achieves \textbf{58.3\%} LLM score, surpassing the \textbf{55.2\%} LLM score of ER-RAG (All in Context), which illustrates the importance of building a source selection system. 

5. Our whole pipeline (ER-RAG (Source Selection)) using an 8B backbone model exceeds any single source and is comparable to (2.9\% boost in LLM score) the best industrial RAG systems, which rely on a complex retrieval pipeline and a large backbone model.

\eat{\stitle{Synthesize Multi-source Dataset.} We present the main result of the Synthesize Multi-source Dataset in Table~\ref{tab:compmix}. This dataset differs from the previous ones because most of the questions require the integration of multiple sources of data. The following conclusions can be drawn:

1. The performance of single-source answers is relatively poor, as multiple sources are required to accurately locate the correct information. After training the model to acquire the ability to handle multi-source data, the performance of the model improves significantly by \textbf{xxx}.

2. Multi-source data can also be effectively routed, enabling the model to better utilize relevant sources and avoid interference from irrelevant ones.}

\begin{figure*}[]
\centering
\begin{tabular}[t]{ccc}
\subfigure[Overall Source Distribution]{
      \psfig{figure=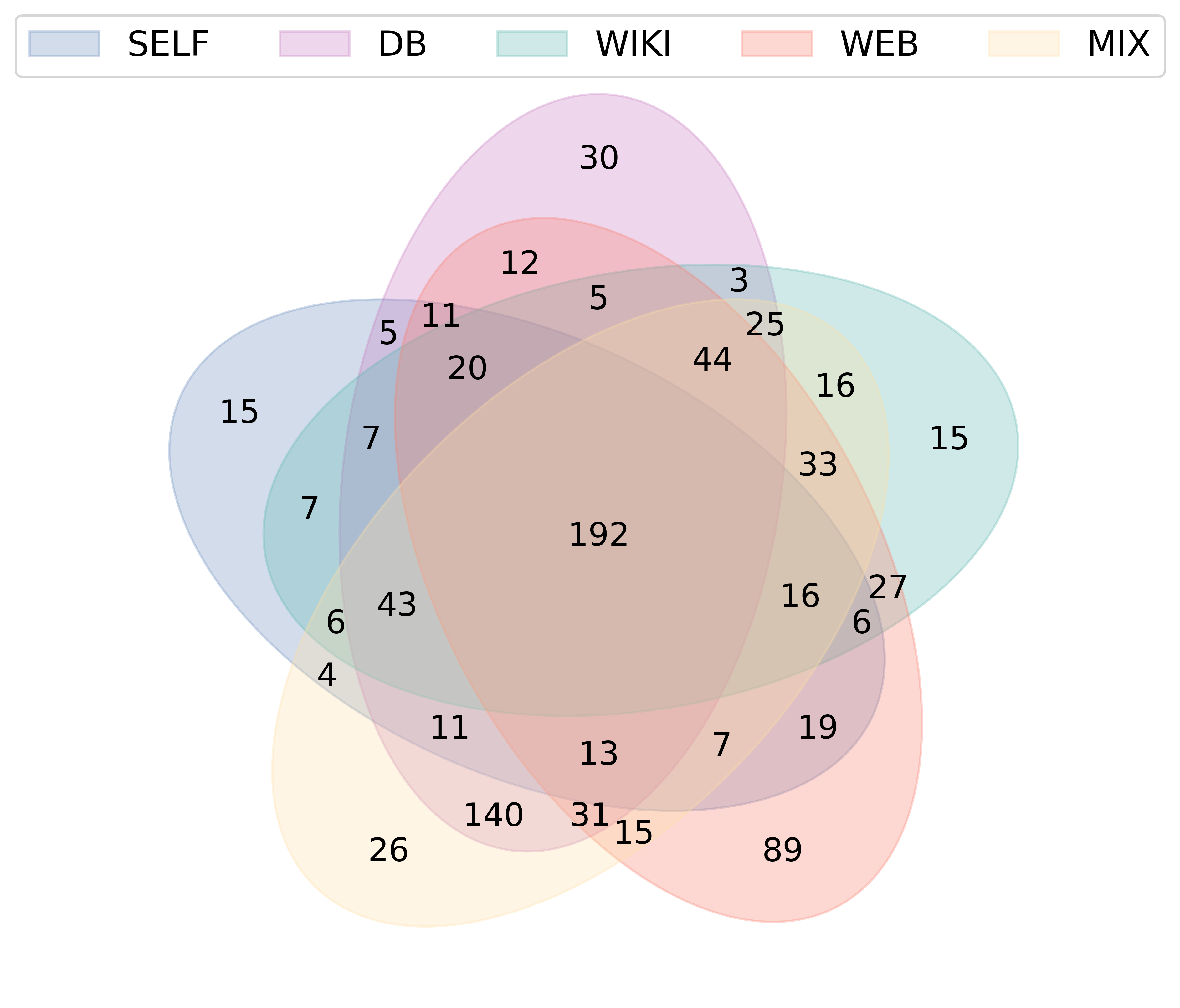, height=4.1cm}
      \label{fig:ab_Web}
      } 
      \hspace{0.5cm}
\subfigure[Cost and Accuracy Comparison on Different Methods]{
        
      \psfig{figure=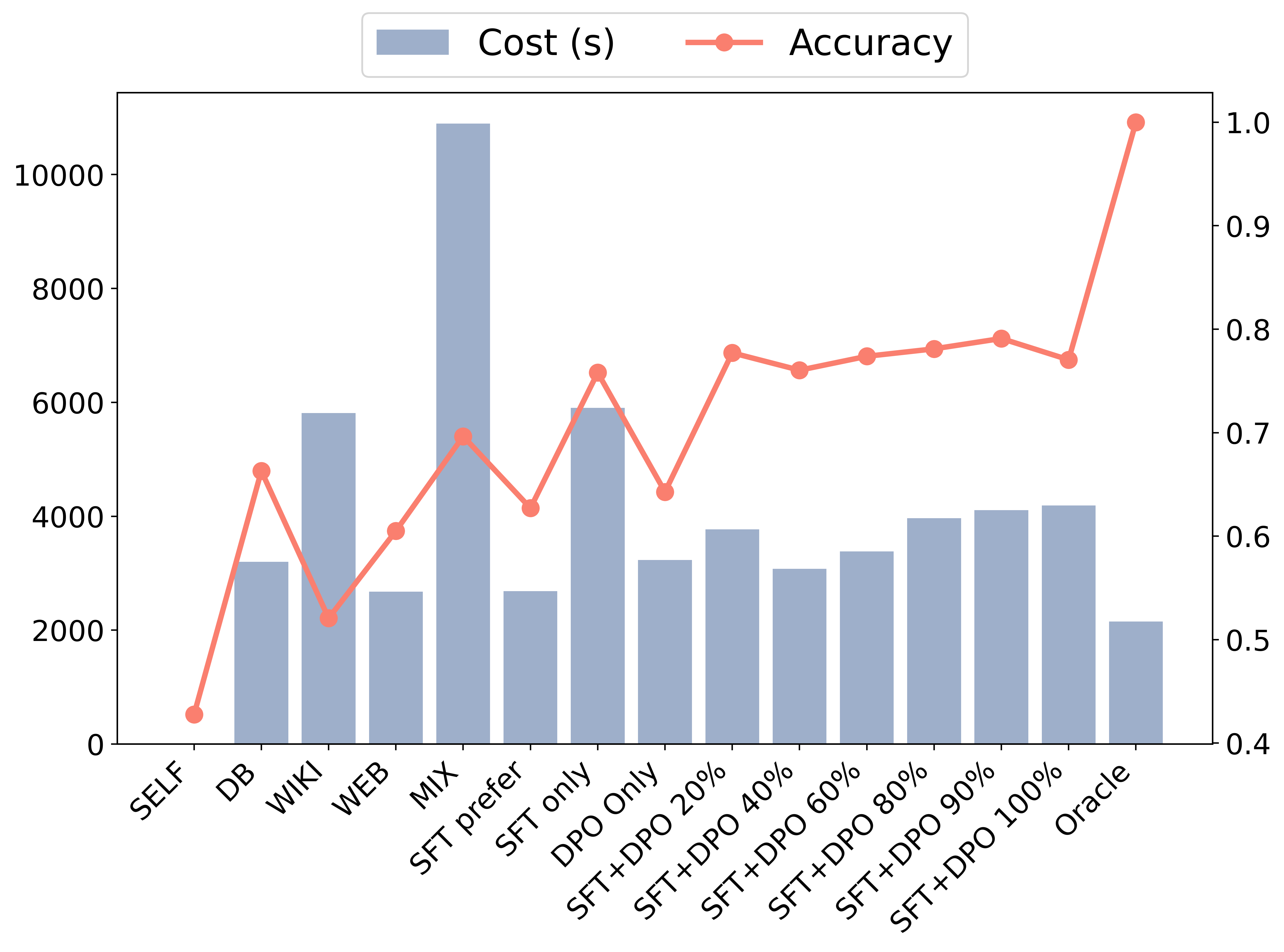, height=4.1cm}
      \label{fig:ab_Web}
      }
      \hspace{0.5cm}
\subfigure[Distribution of Data Sources for Different Methods]{
      \psfig{figure=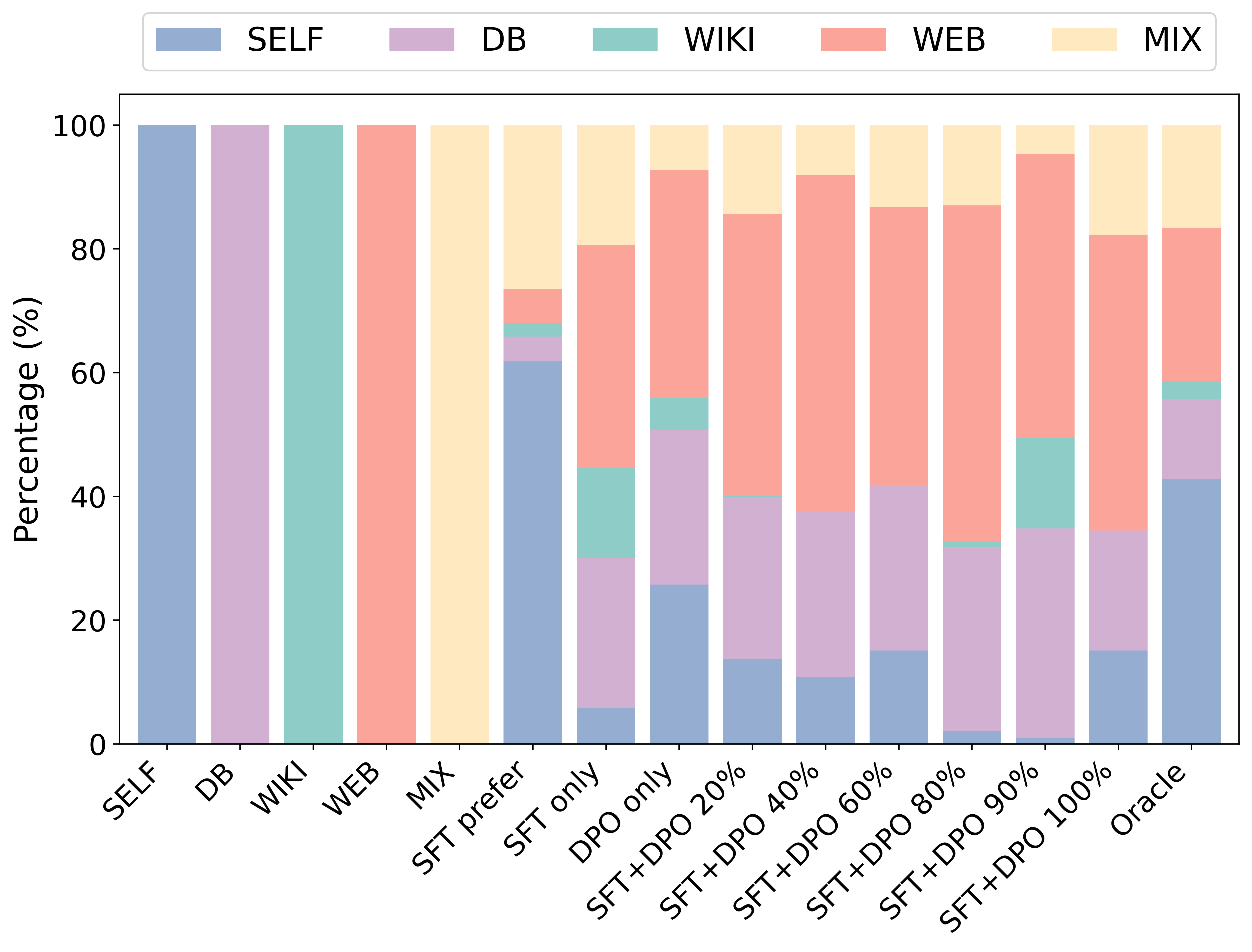, height=4.1cm}
      \label{fig:ab_Web}
      } 
\end{tabular}
\vspace{-0.5cm}
\caption{ Detailed
Source Selection Experiment on CRAG Dataset.}
\label{fig:ab}
\vspace{-0.45cm}
\end{figure*} 
 
 \begin{table}[t]
\centering
\caption{Performance (LLM scores) varying Difference API Training Strategies. (Left: Performance on Synthetic-Test for APIs Trained on Various Source Datasets. Right: MIX-Only vs. MIX-Joint Training on CRAG.)}
\vspace{-0.3cm}
\label{tab:my-table123}
\begin{tabular}{@{}cc@{}} 
\resizebox{0.4\columnwidth}{0.055\textheight}{%
\begin{tabular}{l|c}
\toprule
Source       & Synthetic-Test \\ \midrule
MIX Joint    & 0.380     \\
MIX Only & 0.338     \\
WIKI         & 0.194     \\
DB           & 0.212     \\
SELF         & 0.167     \\ \bottomrule
\end{tabular}%
}
&
\resizebox{0.45\columnwidth}{0.055\textheight}{%
\begin{tabular}{l|cc}
\toprule
Domain       & MIX Joint  & MIX Only \\ \midrule
Movie        & 0.617     & 0.477        \\
Music        & 0.483     & 0.437        \\
Finance      & 0.435     & 0.255        \\
Sports       & 0.454     & 0.218        \\
Open         & 0.568     & 0.476        \\ \midrule
All          & 0.512     & 0.359        \\ \bottomrule
\end{tabular}%
}

 
\end{tabular}
\eat{
\begin{minipage}{0.48\columnwidth}
\centering
(a) Synthetic-Test Performance (LLM score)
\end{minipage}
\begin{minipage}{0.48\columnwidth}
\centering
(b) CRAG Performance (LLM score)
\end{minipage}%
}
\vspace{-0.5cm}
\end{table}

\subsection{Analysis of ER-RAG on Multi-hop Multi-source Queries}

To demonstrate the API generation module's capability on multi-hop multi-source queries, we evaluate it on our synthetic test set. Table~\ref{tab:my-table123} (left) shows the LLM scores, where the API generation module trained on mixed-source APIs improves LLM scores by 18.6\% and 16.8\% compared to using WIKI and DB source agents alone, respectively, highlighting ER-RAG's effectiveness for multi-source multi-hop queries.

As mentioned earlier, it is challenging to build a robust agent with limited training data from a single source. Therefore, we conduct another experiment comparing training only on the synthetic MIX training set (Mix Only) with joint training on the MIX training set and other source training sets (Mix Joint) to demonstrate that API training examples from other sources can significantly benefit API generation on MIX sources. The results of our experiments are presented in Table~\ref{tab:my-table123} (right). From the results, we observe that for API construction on the CRAG dataset, joint training performs approximately 15.3\% better than separate training in terms of overall results. Specifically, for domains with more complex ER extraction operation, such as Finance, joint training significantly outperforms separate training by 18\%. This indicates that joint training can better integrate multi-source data and improve the model's generalization capability in certain challenging scenarios, easing the difficulty of constructing agents.

\begin{table}[]
\centering
\caption{Generalization Test on the CompMix Dataset (Colored numbers indicate the best results among a certain type of methods, ER-RAG trained on CRAG Dataset).}
\vspace{-0.2cm}
\label{tab:compmix}
\resizebox{\columnwidth}{!}{%
\begin{tabular}{c|l|llll}
\toprule
\multirow{2}{*}{Types} & \multicolumn{1}{c|}{\multirow{2}{*}{Methods}} & \multicolumn{4}{c}{LLM scores}\\ \cline{3-6} 
 & \multicolumn{1}{c|}{} & Movie & Music & Sports& All \\ \midrule
\multirow{4}{*}{\begin{tabular}[c]{@{}c@{}}LLM \\ only\end{tabular}} & Llama3 8B & 60.4\ ($\downarrow$19.0)& 56.0\ ($\downarrow$20.5)& 55.3\ ($\downarrow$21.4)& 57.2\ ($\downarrow$20.1)\\
 & GPT-4o-mini& 71.3\ ($\downarrow$8.1)& 66.4\ ($\downarrow$10.1)& 66.5\ ($\downarrow$10.2)& 68.1\ ($\downarrow$9.2)\\
 & GPT-4o & \textcolor{blue}{\textbf{80.8}\ ($\uparrow$1.4)} & \textcolor{blue}{\textbf{75.8}\ ($\downarrow$0.7)} & \textcolor{blue}{\textbf{75.9}\ ($\downarrow$0.8)} & \textcolor{blue}{\textbf{77.5}\ ($\uparrow$0.2)} \\
 & DeepSeek v3& 74.8\ ($\downarrow$4.6)& 74.7\ ($\downarrow$1.8)& 75.5\ ($\downarrow$1.2)& 75.0\ ($\downarrow$2.3)\\ \midrule
\multirow{3}{*}{\begin{tabular}[c]{@{}c@{}}Industry \\ systems\end{tabular}} & ChatGPT Plus (GPT-4o) & 78.8\ ($\downarrow$0.6)& 78.1\ ($\uparrow$1.6)& 76.2\ ($\downarrow$0.5)& 77.7\ ($\uparrow$0.4)\\
 & Claude-3-5-sonnet Pro & 79.3\ ($\downarrow$0.1)& 76.5\ ($\downarrow$0.0)& 74.8\ ($\downarrow$1.9)& 76.8\ ($\downarrow$0.5)\\
 & Perplexity (405B) & \textcolor{violet}{\textbf{80.0}\ ($\uparrow$0.6)} & \textcolor{violet}{\textbf{78.3}\ ($\uparrow$1.8)}& \textcolor{violet}{\textbf{77.7}\ ($\uparrow$1.0)} & \textcolor{violet}{\textbf{78.7}\ ($\uparrow$1.4)} \\ \hline
\multirow{5}{*}{\begin{tabular}[c]{@{}c@{}}Single \\ data source\\ (Llama3 8B)\end{tabular}} & SuperSQL (DB, GPT-4o)& 60.4\ ($\downarrow$19.0)& 54.2\ ($\downarrow$22.3)& 55.0\ ($\downarrow$21.7)& 56.5\ ($\downarrow$20.8)\\
 & ER-RAG (DB, GPT-4o)& 60.7\ ($\downarrow$18.7)& 49.5\ ($\downarrow$27.0)& 56.6\ ($\downarrow$20.1)& 55.7\ ($\downarrow$21.6)\\
 & ER-RAG (WEB Chunk)& \textcolor{orange}{\textbf{76.6}\ ($\downarrow$2.8)} & \textcolor{orange}{\textbf{76.5}\ ($\downarrow$0.0)} & \textcolor{orange}{\textbf{76.6}\ ($\downarrow$0.1)} & \textcolor{orange}{\textbf{76.5}\ ($\downarrow$0.8)} \\
 & ER-RAG (WEB ER) & 75.9\ ($\downarrow$3.5)& 73.6\ ($\downarrow$2.9)& 72.4\ ($\downarrow$4.3)& 74.0\ ($\downarrow$3.3)\\
 & ER-RAG (WIKI ER)& 75.4\ ($\downarrow$4.0)& 67.9\ ($\downarrow$8.6)& 67.2\ ($\downarrow$9.5)& 69.9\ ($\downarrow$7.4)\\ \midrule
\multirow{4}{*}{\begin{tabular}[c]{@{}c@{}}Hybrid \\ data sources\\ (Llama3 8B)\end{tabular}}  
 & ER-RAG (MIX)& 76.2\ ($\downarrow$3.2)& 67.9\ ($\downarrow$8.6)& 66.8\ ($\downarrow$9.9)& 70.2\ ($\downarrow$7.1)\\
 & ER-RAG (All in Context)& 76.2\ ($\downarrow$3.2)& 72.8\ ($\downarrow$3.7)& 72.1\ ($\downarrow$4.6)& 73.6\ ($\downarrow$3.7)\\
 & ER-RAG (Source Selection)& \textcolor{mygreen!110}{\textbf{79.4}} & \textcolor{mygreen!110}{\textbf{76.5}} & \textcolor{mygreen!110}{\textbf{76.7}} & \textcolor{mygreen!110}{\textbf{77.3}} \\
 & \textcolor{gray}{ER-RAG (Oracle)} & \textcolor{gray}{87.0}& \textcolor{gray}{86.6}& \textcolor{gray}{86.4}&\textcolor{gray}{86.7}\\ \bottomrule
\end{tabular}%
}
\vspace{-0.5cm}
\end{table}

\subsection{Generalization of ER-RAG}
 
To demonstrate the generalization of ER-RAG, we apply ER-RAG trained on the CRAG dataset directly on the CompMix dataset, and the results are in Table~\ref{tab:compmix}. Similar conclusions can be made about the effect of RAG and source selection. The following conclusions can be made: 1. The DB source from the CRAG dataset doesn't perform well on the CompMix dataset, illustrating that the data source should be properly selected to fit in the queries. Irrelevant information may harm the RAG performance. 2. The agent trained on the CRAG training set transfers well to the CompMix dataset. The combined heterogeneous sources results outperform single sources and achieve similar results compared with the best industrial result, reaching 89.2\% of all the potential of the RAG system. The source selection module eliminate the poor results from the DB source, relying more heavily on the WEB and WIKI sources.

\subsection{Source Selection  Study}
In this subsection, we focus on source selection. Figure~\ref{fig:ab}(a) shows the source distribution, Figure \ref{fig:ab}(b) presents efficiency and accuracy of sources and methods, and Figure~\ref{fig:ab}(c) displays the data source distribution.
 We use \textbf{SELF} (no RAG), \textbf{DB}, \textbf{WIKI}, \textbf{WEB}, and \textbf{MIX} (Multi-source API) to represent different sources. \textbf{DB} includes various databases (\eg Movie\_DB), and \textbf{MIX} combines hybrid sources (\eg Movie\_DB + WIKI).  We use \textbf{DB} and \textbf{MIX}  for simplicity.

We define training strategies for source selection:   \textbf{SFT prefer}  uses the correct source with minimal execution time for supervised fine-tuning (SFT), and \textbf{SFT only} uses all correct sources as labels. \textbf{DPO only}  fine-tunes using execution time preference pairs. \textbf{SFT+DPO x\%}  combines SFT to identify all correct sources and DPO with x\% of preference pairs. \textbf{Oracle} represents the ideal scenario of selecting the correct source with minimal execution time. Metrics include end-to-end retrieval time (in seconds, 0 for SELF), and the accuracy compared to the oracle situation (1 for Oracle).

\stitle{Overall Source Distribution.} As shown in Figure~\ref{fig:ab}(a), each source has unique queries in the CRAG test dataset that only it can answer, highlighting the importance of integrating sources to enhance the RAG system's potential. Additionally, many queries can be answered by multiple sources, making it crucial to select the correct source with minimal retrieval time.

\stitle{Cost and Accuracy Comparison on Different Methods.} We can draw two conclusions from Figure~~\ref{fig:ab}(b). 1. The execution time for different sources varies significantly, ranging from 0s for no RAG to $10^4s$ for multi-source APIs. For single sources, WIKI costs the most time for retrieval as the pipeline requires much running time for entity selection and attribute extraction. 2. The \textbf{SFT prefer}, \textbf{SFT only}, and \textbf{DPO only} method all have their own drawbacks. Although the \textbf{SFT prefer} approach is the fastest among the tuning methods, the accuracy deteriorates a lot. The \textbf{SFT only} method can choose the source with higher accuracy, while due to the lack of perception of the execution time, the overall retrieval time cost is relatively high. The \textbf{DPO only} method tends to choose the source with a lower cost, while the accuracy is sacrificed. Our two-stage training method can achieve a balance between accuracy and efficiency. As we can see \textbf{SFT+DPO} can achieve slightly higher accuracy compared with \textbf{SFT only}, while the time cost is significantly lower than the \textbf{SFT only} result. Compared with the \textbf{DPO only} result, the accuracy result is much better.

\stitle{Distribution of Different Methods Across Data Sources.} We can draw more insights from the chosen source distribution. From the oracle perspective, choosing a large proportion of SELF source (No RAG) can achieve good results. However, it's hard for the LLM to determine whether a question can be answered without RAG, as the \textbf{DPO only} and \textbf{SFT prefer} results choosing the SELF sources more frequently in performance loss. Compared with the \textbf{SFT only} method, our proposed method cautiously chooses more SELF source to ensure higher efficiency and maintain  accuracy.
\subsection{Further Ablation Study on ER-RAG}
\begin{figure}[t]
\centering
\label{fig:ab}
\begin{tabular}[t]{ccc}
  \vspace{-0.3cm}
\subfigure[DB ]{
      \psfig{figure=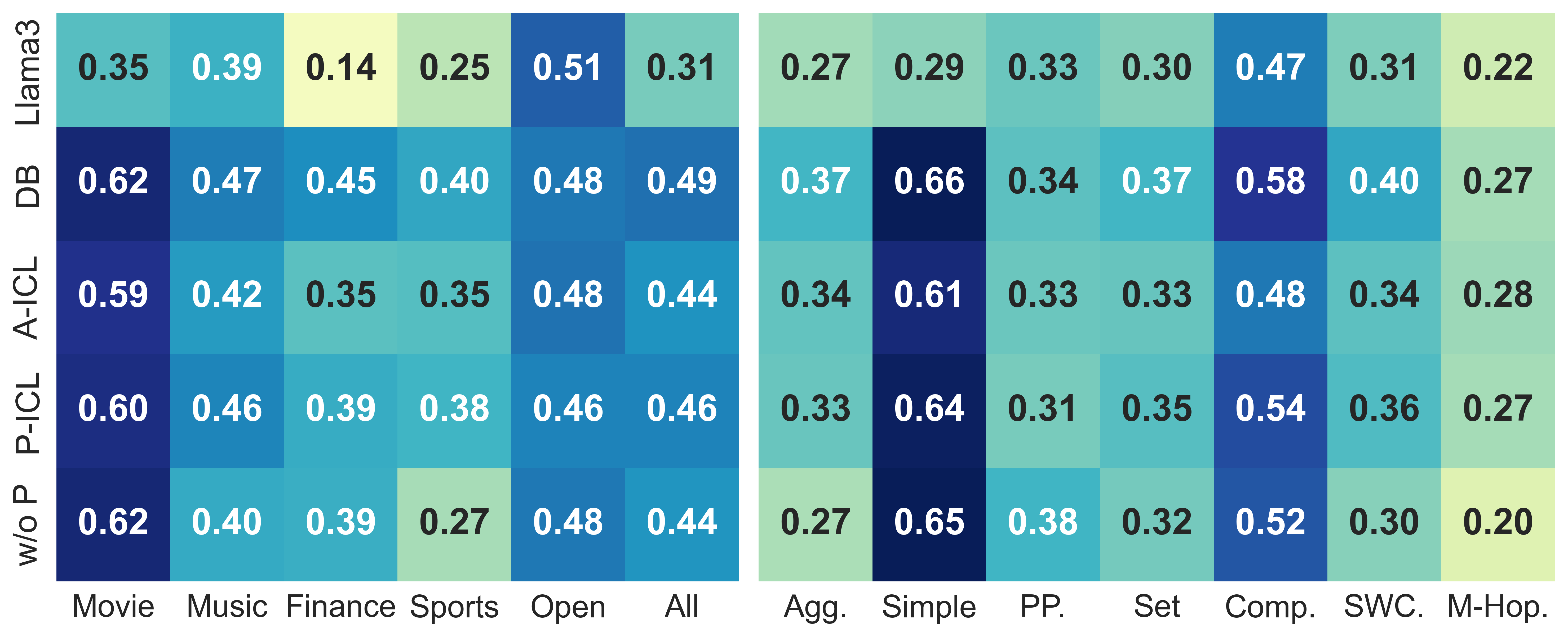,width = 0.9\linewidth}
      \label{fig:ab_db}
      }
      \\
\vspace{-0.3cm}
\subfigure[WIKI ]{
      \psfig{figure=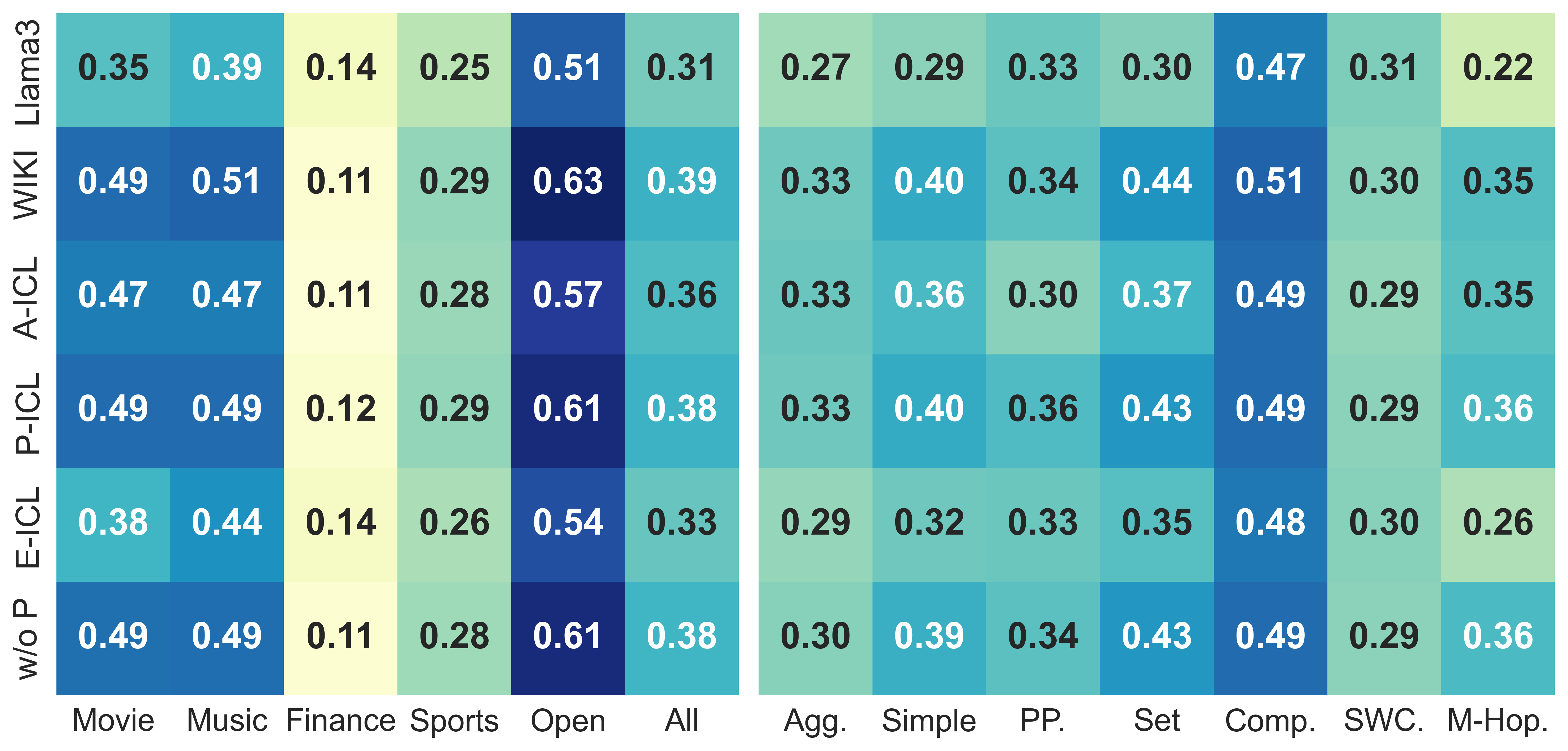,width = 0.9\linewidth}
      \label{fig:ab_Wiki}
      }
      \\
\vspace{-0.3cm}
\subfigure[WEB]{
\psfig{figure=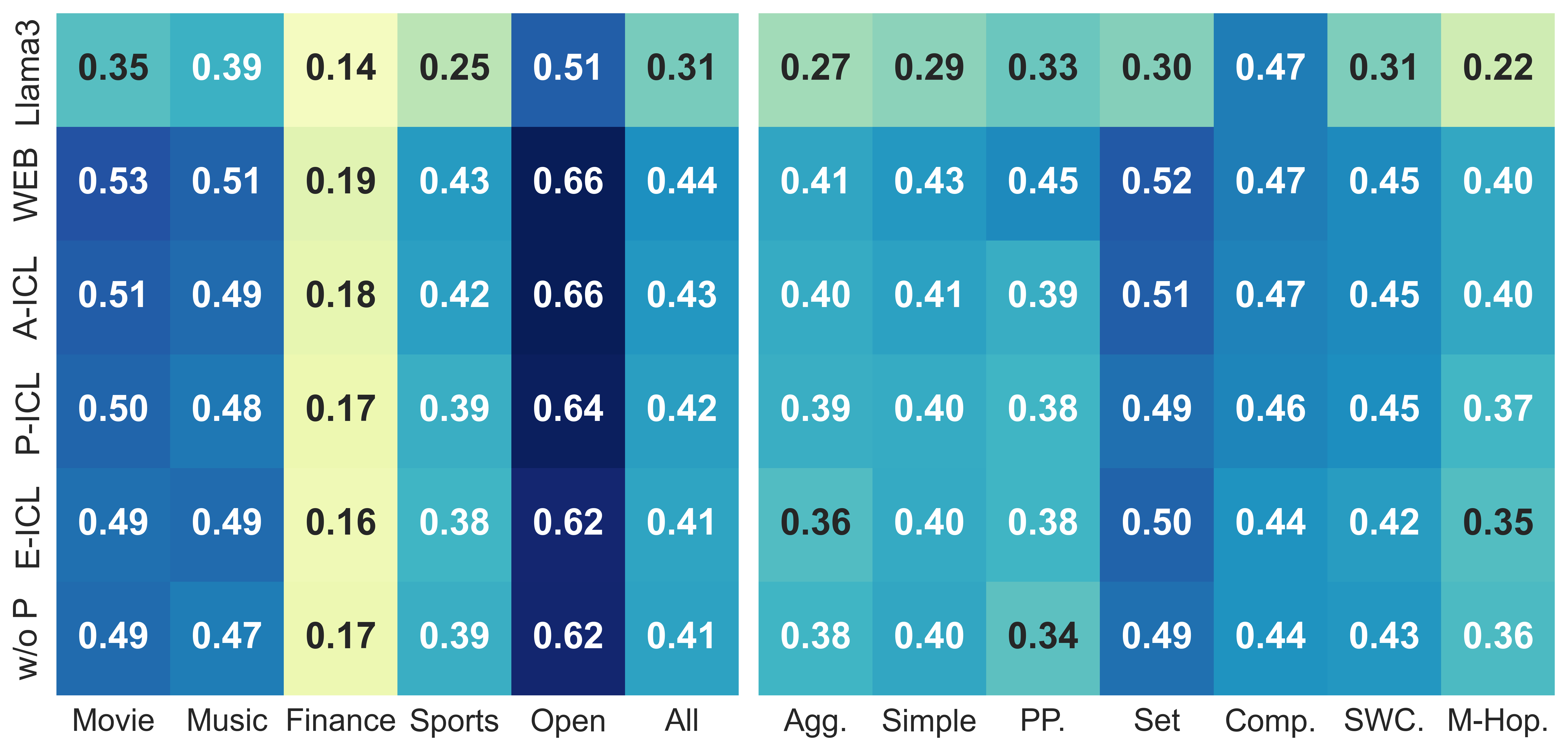,width =0.9\linewidth}
\label{fig:ab_Web}
}
\end{tabular}
\vspace{-0.3cm}
\caption{ Detailed
Ablation Experiment on CRAG Dataset: Performance Across Domains and Question Types. Question types include Aggregation (Agg.), Post-Processing (PP.), Set, Comparison (Comp.), Simple with Condition (SWC.), and Multi-Hop (M-Hop).}
\label{fig:ab}
\vspace{-0.4cm}
\end{figure} 
In this subsection, we perform an ablation study to assess the impact of different ER-RAG modules across domains and question types on CRAG (Figure~\ref{fig:ab}). The vertical axis shows domain and question type results, while the horizontal axis represents methods: \textbf{DB}, \textbf{WIKI}, \textbf{WEB} (complete model in each source) and \textbf{Llama3} (base model without RAG).

Other methods can be grouped into two categories. The first category replaces fine-tuning with in-context learning (ICL) for different components: \textbf{API\_ICL (A-ICL)} represents using ICL to generate API calls, \textbf{Post-processing\_ICL (P-ICL)} denotes using ICL to produce post-processing results, and \textbf{Entity\_ICL (E-ICL)} applies ICL for information extraction in WIKI entity disambiguation and $GET$/$JOIN$ attribute modules. The second category involves removing specific modules entirely: \textbf{Without Post-processing (w/o P)} refers to eliminating the post-processing module.

\eat{
\begin{itemize}
    \item \textbf{API\_ICL} (A-ICL in DB, WIKI and WEB chart): Using in-context learning (ICL) to generate API calls without fine-tuning.
    \item \textbf{Tool\_ICL} (T-ICL in DB, WIKI and WEB Chart): Using ICL to generate post-processing results without fine-tuning.
    \item \textbf{Entity\_ICL} (E-ICL in WIKI and WEB Chart): Present only in the WIKI entity disambiguation and $GET$ and $JOIN$'s attribute extraction module, which uses ICL to extract information instead of fine-tuning.
    \item \textbf{Without Tool} (w/o T in DB, WIKI and WEB Chart): Removing the tool module entirely.
\end{itemize}
}
From the results, we can draw the following conclusions:

1. Across all modules, models that utilize fine-tuning outperform those relying solely on ICL. This demonstrates the importance of leveraging strong models for APIs, fine-tuning source selection results based on execution outcomes, and distilling knowledge for post-processing and entities from larger and stronger LLMs. The results suggest that the 8B base model is insufficient when only using ICL. Significant improvements can be distilled from large models such as DeepSeek/GPT-4 and manually annotated datasets.

2. The post-processing module plays a crucial role, particularly in the DB source, where it achieves a performance gain of \textbf{5\%}. From the perspective of question types, the performance boost is most significant on the simple with conditions (\textbf{12\%}) question type, as the conditions in these queries require more intensive post-processing. This highlights this module's ability to address computations required in database queries. However, for other sources where retrieval is inherently less precise, the contribution of the post-processing module is relatively limited.

\section{Conclusion and Future Work}
\label{sec-conclusion}

Motivated by the need for evidence collection from heterogeneous data sources in RAG, this paper introduces ER-RAG, which enhances LLM capabilities using ER-based unified APIs for HTML, knowledge graphs, and relational databases. ER-RAG enables seamless interaction between data sources, provides diverse training instances for API generation, and balances response time and performance through DPO-based source selection. Experiments demonstrate that ER-RAG matches source-specific or commercial RAG performance while excelling on multi-source tasks.

ER-RAG can be extended in the following interesting directions. First, ER-RAG can leverage more advanced retrieval techniques proposed for the specific source. Such an improvement can be done at the implementation of the unified Access APIs and will not impact the entire process. Second, ER-RAG is expected to combine more sources, like multi-modal data sources, into the RAG pipeline. Third, ER-RAG will collect the user's feedback and incorporate it in fine-tuning the LLM agents timely so as to improve the user's experience.

\clearpage

\bibliographystyle{ACM-Reference-Format}
\bibliography{main}


\begin{thebibliography}{48}


\ifx \showCODEN    \undefined \def \showCODEN     #1{\unskip}     \fi
\ifx \showDOI      \undefined \def \showDOI       #1{#1}\fi
\ifx \showISBNx    \undefined \def \showISBNx     #1{\unskip}     \fi
\ifx \showISBNxiii \undefined \def \showISBNxiii  #1{\unskip}     \fi
\ifx \showISSN     \undefined \def \showISSN      #1{\unskip}     \fi
\ifx \showLCCN     \undefined \def \showLCCN      #1{\unskip}     \fi
\ifx \shownote     \undefined \def \shownote      #1{#1}          \fi
\ifx \showarticletitle \undefined \def \showarticletitle #1{#1}   \fi
\ifx \showURL      \undefined \def \showURL       {\relax}        \fi
\providecommand\bibfield[2]{#2}
\providecommand\bibinfo[2]{#2}
\providecommand\natexlab[1]{#1}
\providecommand\showeprint[2][]{arXiv:#2}

\bibitem[\protect\citeauthoryear{Asai, Wu, Wang, Sil, and Hajishirzi}{Asai et~al\mbox{.}}{2024}]%
        {selfrag}
\bibfield{author}{\bibinfo{person}{Akari Asai}, \bibinfo{person}{Zeqiu Wu}, \bibinfo{person}{Yizhong Wang}, \bibinfo{person}{Avirup Sil}, {and} \bibinfo{person}{Hannaneh Hajishirzi}.} \bibinfo{year}{2024}\natexlab{}.
\newblock \showarticletitle{Self-RAG: Learning to Retrieve, Generate, and Critique through Self-Reflection}. In \bibinfo{booktitle}{\emph{ICLR}}.
\newblock


\bibitem[\protect\citeauthoryear{Bai, Wang, Li, Ding, Zhang, and Gao}{Bai et~al\mbox{.}}{2021}]%
        {atjnet}
\bibfield{author}{\bibinfo{person}{Jinze Bai}, \bibinfo{person}{Jialin Wang}, \bibinfo{person}{Zhao Li}, \bibinfo{person}{Donghui Ding}, \bibinfo{person}{Ji Zhang}, {and} \bibinfo{person}{Jun Gao}.} \bibinfo{year}{2021}\natexlab{}.
\newblock \showarticletitle{ATJ-Net: Auto-Table-Join Network for Automatic Learning on Relational Databases}. In \bibinfo{booktitle}{\emph{{WWW}}}. \bibinfo{pages}{1540--1551}.
\newblock


\bibitem[\protect\citeauthoryear{Biswal, Patel, Jha, Kamsetty, Liu, Gonzalez, Guestrin, and Zaharia}{Biswal et~al\mbox{.}}{2024}]%
        {tag}
\bibfield{author}{\bibinfo{person}{Asim Biswal}, \bibinfo{person}{Liana Patel}, \bibinfo{person}{Siddarth Jha}, \bibinfo{person}{Amog Kamsetty}, \bibinfo{person}{Shu Liu}, \bibinfo{person}{Joseph~E. Gonzalez}, \bibinfo{person}{Carlos Guestrin}, {and} \bibinfo{person}{Matei Zaharia}.} \bibinfo{year}{2024}\natexlab{}.
\newblock \showarticletitle{Text2SQL is Not Enough: Unifying {AI} and Databases with {TAG}}.
\newblock \bibinfo{journal}{\emph{CoRR}}  \bibinfo{volume}{abs/2408.14717} (\bibinfo{year}{2024}).
\newblock


\bibitem[\protect\citeauthoryear{Bourhis, Reutter, Su{\'{a}}rez, and Vrgoc}{Bourhis et~al\mbox{.}}{2017}]%
        {json}
\bibfield{author}{\bibinfo{person}{Pierre Bourhis}, \bibinfo{person}{Juan~L. Reutter}, \bibinfo{person}{Fernando Su{\'{a}}rez}, {and} \bibinfo{person}{Domagoj Vrgoc}.} \bibinfo{year}{2017}\natexlab{}.
\newblock \showarticletitle{{JSON:} Data model, Query languages and Schema specification}. In \bibinfo{booktitle}{\emph{PODS}}.
\newblock


\bibitem[\protect\citeauthoryear{Brown, Mann, Ryder, Subbiah, Kaplan, Dhariwal, Neelakantan, Shyam, Sastry, Askell, Agarwal, Herbert{-}Voss, Krueger, Henighan, Child, Ramesh, Ziegler, Wu, Winter, Hesse, Chen, Sigler, Litwin, Gray, Chess, Clark, Berner, McCandlish, Radford, Sutskever, and Amodei}{Brown et~al\mbox{.}}{2020}]%
        {LLMfewshot}
\bibfield{author}{\bibinfo{person}{Tom~B. Brown}, \bibinfo{person}{Benjamin Mann}, \bibinfo{person}{Nick Ryder}, \bibinfo{person}{Melanie Subbiah}, \bibinfo{person}{Jared Kaplan}, \bibinfo{person}{Prafulla Dhariwal}, \bibinfo{person}{Arvind Neelakantan}, \bibinfo{person}{Pranav Shyam}, \bibinfo{person}{Girish Sastry}, \bibinfo{person}{Amanda Askell}, \bibinfo{person}{Sandhini Agarwal}, \bibinfo{person}{Ariel Herbert{-}Voss}, \bibinfo{person}{Gretchen Krueger}, \bibinfo{person}{Tom Henighan}, \bibinfo{person}{Rewon Child}, \bibinfo{person}{Aditya Ramesh}, \bibinfo{person}{Daniel~M. Ziegler}, \bibinfo{person}{Jeffrey Wu}, \bibinfo{person}{Clemens Winter}, \bibinfo{person}{Christopher Hesse}, \bibinfo{person}{Mark Chen}, \bibinfo{person}{Eric Sigler}, \bibinfo{person}{Mateusz Litwin}, \bibinfo{person}{Scott Gray}, \bibinfo{person}{Benjamin Chess}, \bibinfo{person}{Jack Clark}, \bibinfo{person}{Christopher Berner}, \bibinfo{person}{Sam McCandlish}, \bibinfo{person}{Alec Radford}, \bibinfo{person}{Ilya Sutskever},
  {and} \bibinfo{person}{Dario Amodei}.} \bibinfo{year}{2020}\natexlab{}.
\newblock \showarticletitle{Language Models are Few-Shot Learners}. In \bibinfo{booktitle}{\emph{NeurIPS 2020}}.
\newblock


\bibitem[\protect\citeauthoryear{Cappuzzo, Papotti, and Thirumuruganathan}{Cappuzzo et~al\mbox{.}}{2020}]%
        {embdi}
\bibfield{author}{\bibinfo{person}{Riccardo Cappuzzo}, \bibinfo{person}{Paolo Papotti}, {and} \bibinfo{person}{Saravanan Thirumuruganathan}.} \bibinfo{year}{2020}\natexlab{}.
\newblock \showarticletitle{Creating Embeddings of Heterogeneous Relational Datasets for Data Integration Tasks}. In \bibinfo{booktitle}{\emph{SIGMOD}}. \bibinfo{pages}{1335--1349}.
\newblock


\bibitem[\protect\citeauthoryear{Chen, Stanovsky, Singh, and Gardner}{Chen et~al\mbox{.}}{2019}]%
        {chen2019evaluating}
\bibfield{author}{\bibinfo{person}{Anthony Chen}, \bibinfo{person}{Gabriel Stanovsky}, \bibinfo{person}{Sameer Singh}, {and} \bibinfo{person}{Matt Gardner}.} \bibinfo{year}{2019}\natexlab{}.
\newblock \showarticletitle{Evaluating question answering evaluation}. In \bibinfo{booktitle}{\emph{Proceedings of the 2nd workshop on machine reading for question answering}}. \bibinfo{pages}{119--124}.
\newblock


\bibitem[\protect\citeauthoryear{Chen, Xiao, Zhang, Luo, Lian, and Liu}{Chen et~al\mbox{.}}{2024a}]%
        {chen2024bge}
\bibfield{author}{\bibinfo{person}{Jianlv Chen}, \bibinfo{person}{Shitao Xiao}, \bibinfo{person}{Peitian Zhang}, \bibinfo{person}{Kun Luo}, \bibinfo{person}{Defu Lian}, {and} \bibinfo{person}{Zheng Liu}.} \bibinfo{year}{2024}\natexlab{a}.
\newblock \showarticletitle{Bge m3-embedding: Multi-lingual, multi-functionality, multi-granularity text embeddings through self-knowledge distillation}.
\newblock \bibinfo{journal}{\emph{arXiv preprint arXiv:2402.03216}} (\bibinfo{year}{2024}).
\newblock


\bibitem[\protect\citeauthoryear{Chen, Zhang, and Roth}{Chen et~al\mbox{.}}{2024b}]%
        {multabret}
\bibfield{author}{\bibinfo{person}{Peter~Baile Chen}, \bibinfo{person}{Yi Zhang}, {and} \bibinfo{person}{Dan Roth}.} \bibinfo{year}{2024}\natexlab{b}.
\newblock \showarticletitle{Is Table Retrieval a Solved Problem? Exploring Join-Aware Multi-Table Retrieval}. In \bibinfo{booktitle}{\emph{ACL}}. \bibinfo{pages}{2687--2699}.
\newblock


\bibitem[\protect\citeauthoryear{Chen}{Chen}{1975}]%
        {ermodel}
\bibfield{author}{\bibinfo{person}{Peter~P. Chen}.} \bibinfo{year}{1975}\natexlab{}.
\newblock \showarticletitle{The Entity-Relationship Model: Toward a Unified View of Data}. In \bibinfo{booktitle}{\emph{Proceedings of VLDB, 1975}}. \bibinfo{pages}{173}.
\newblock


\bibitem[\protect\citeauthoryear{Christiano, Leike, Brown, Martic, Legg, and Amodei}{Christiano et~al\mbox{.}}{2017}]%
        {RLHF}
\bibfield{author}{\bibinfo{person}{Paul~F. Christiano}, \bibinfo{person}{Jan Leike}, \bibinfo{person}{Tom~B. Brown}, \bibinfo{person}{Miljan Martic}, \bibinfo{person}{Shane Legg}, {and} \bibinfo{person}{Dario Amodei}.} \bibinfo{year}{2017}\natexlab{}.
\newblock \showarticletitle{Deep Reinforcement Learning from Human Preferences}. In \bibinfo{booktitle}{\emph{NIPS}}. \bibinfo{pages}{4299--4307}.
\newblock


\bibitem[\protect\citeauthoryear{Christmann, Roy, and Weikum}{Christmann et~al\mbox{.}}{2024}]%
        {CompMix}
\bibfield{author}{\bibinfo{person}{Philipp Christmann}, \bibinfo{person}{Rishiraj~Saha Roy}, {and} \bibinfo{person}{Gerhard Weikum}.} \bibinfo{year}{2024}\natexlab{}.
\newblock \showarticletitle{CompMix: {A} Benchmark for Heterogeneous Question Answering}. In \bibinfo{booktitle}{\emph{WWW}}. \bibinfo{pages}{1091--1094}.
\newblock


\bibitem[\protect\citeauthoryear{Devlin, Chang, Lee, and Toutanova}{Devlin et~al\mbox{.}}{2019}]%
        {bert}
\bibfield{author}{\bibinfo{person}{Jacob Devlin}, \bibinfo{person}{Ming{-}Wei Chang}, \bibinfo{person}{Kenton Lee}, {and} \bibinfo{person}{Kristina Toutanova}.} \bibinfo{year}{2019}\natexlab{}.
\newblock \showarticletitle{{BERT:} Pre-training of Deep Bidirectional Transformers for Language Understanding}. In \bibinfo{booktitle}{\emph{NAACL-HLT 2019}}. \bibinfo{pages}{4171--4186}.
\newblock


\bibitem[\protect\citeauthoryear{Dong, Li, Dai, Zheng, Ma, Li, Xia, Xu, Wu, Chang, et~al\mbox{.}}{Dong et~al\mbox{.}}{2024}]%
        {icl_dong2024survey}
\bibfield{author}{\bibinfo{person}{Qingxiu Dong}, \bibinfo{person}{Lei Li}, \bibinfo{person}{Damai Dai}, \bibinfo{person}{Ce Zheng}, \bibinfo{person}{Jingyuan Ma}, \bibinfo{person}{Rui Li}, \bibinfo{person}{Heming Xia}, \bibinfo{person}{Jingjing Xu}, \bibinfo{person}{Zhiyong Wu}, \bibinfo{person}{Baobao Chang}, {et~al\mbox{.}}} \bibinfo{year}{2024}\natexlab{}.
\newblock \showarticletitle{A Survey on In-context Learning}. In \bibinfo{booktitle}{\emph{Proceedings of the 2024 Conference on Empirical Methods in Natural Language Processing}}. \bibinfo{pages}{1107--1128}.
\newblock


\bibitem[\protect\citeauthoryear{Dong}{Dong}{2023}]%
        {kg_Dong2023}
\bibfield{author}{\bibinfo{person}{Xin~Luna Dong}.} \bibinfo{year}{2023}\natexlab{}.
\newblock \showarticletitle{Generations of Knowledge Graphs: The Crazy Ideas and the Business Impact}.
\newblock \bibinfo{journal}{\emph{Proc. {VLDB} Endow.}} \bibinfo{volume}{16}, \bibinfo{number}{12} (\bibinfo{year}{2023}), \bibinfo{pages}{4130--4137}.
\newblock


\bibitem[\protect\citeauthoryear{Dong, Gabrilovich, Murphy, Dang, Horn, Lugaresi, Sun, and Zhang}{Dong et~al\mbox{.}}{2015}]%
        {Trust_source2015}
\bibfield{author}{\bibinfo{person}{Xin~Luna Dong}, \bibinfo{person}{Evgeniy Gabrilovich}, \bibinfo{person}{Kevin Murphy}, \bibinfo{person}{Van Dang}, \bibinfo{person}{Wilko Horn}, \bibinfo{person}{Camillo Lugaresi}, \bibinfo{person}{Shaohua Sun}, {and} \bibinfo{person}{Wei Zhang}.} \bibinfo{year}{2015}\natexlab{}.
\newblock \showarticletitle{Knowledge-Based Trust: Estimating the Trustworthiness of Web Sources}.
\newblock \bibinfo{journal}{\emph{Proc. {VLDB} Endow.}} \bibinfo{volume}{8}, \bibinfo{number}{9} (\bibinfo{year}{2015}), \bibinfo{pages}{938--949}.
\newblock


\bibitem[\protect\citeauthoryear{Dong, Saha, and Srivastava}{Dong et~al\mbox{.}}{2012}]%
        {less_Dong2012}
\bibfield{author}{\bibinfo{person}{Xin~Luna Dong}, \bibinfo{person}{Barna Saha}, {and} \bibinfo{person}{Divesh Srivastava}.} \bibinfo{year}{2012}\natexlab{}.
\newblock \showarticletitle{Less is More: Selecting Sources Wisely for Integration}.
\newblock \bibinfo{journal}{\emph{Proc. {VLDB} Endow.}} \bibinfo{volume}{6}, \bibinfo{number}{2} (\bibinfo{year}{2012}), \bibinfo{pages}{37--48}.
\newblock


\bibitem[\protect\citeauthoryear{Fan, Ding, Ning, Wang, Li, Yin, Chua, and Li}{Fan et~al\mbox{.}}{2024}]%
        {ragsurvey24}
\bibfield{author}{\bibinfo{person}{Wenqi Fan}, \bibinfo{person}{Yujuan Ding}, \bibinfo{person}{Liangbo Ning}, \bibinfo{person}{Shijie Wang}, \bibinfo{person}{Hengyun Li}, \bibinfo{person}{Dawei Yin}, \bibinfo{person}{Tat-Seng Chua}, {and} \bibinfo{person}{Qing Li}.} \bibinfo{year}{2024}\natexlab{}.
\newblock \showarticletitle{A Survey on RAG Meeting LLMs: Towards Retrieval-Augmented Large Language Models}. In \bibinfo{booktitle}{\emph{Proceedings of SIGKDD 2024}}. \bibinfo{pages}{6491–6501}.
\newblock


\bibitem[\protect\citeauthoryear{Gao, Wang, Li, Sun, Qian, Ding, and Zhou}{Gao et~al\mbox{.}}{2024}]%
        {gao2024text}
\bibfield{author}{\bibinfo{person}{Dawei Gao}, \bibinfo{person}{Haibin Wang}, \bibinfo{person}{Yaliang Li}, \bibinfo{person}{Xiuyu Sun}, \bibinfo{person}{Yichen Qian}, \bibinfo{person}{Bolin Ding}, {and} \bibinfo{person}{Jingren Zhou}.} \bibinfo{year}{2024}\natexlab{}.
\newblock \showarticletitle{Text-to-SQL Empowered by Large Language Models: A Benchmark Evaluation}.
\newblock \bibinfo{journal}{\emph{Proceedings of the VLDB Endowment}} \bibinfo{volume}{17}, \bibinfo{number}{5} (\bibinfo{year}{2024}), \bibinfo{pages}{1132--1145}.
\newblock


\bibitem[\protect\citeauthoryear{Gao, Ma, Lin, and Callan}{Gao et~al\mbox{.}}{2023}]%
        {gao2023precise}
\bibfield{author}{\bibinfo{person}{Luyu Gao}, \bibinfo{person}{Xueguang Ma}, \bibinfo{person}{Jimmy Lin}, {and} \bibinfo{person}{Jamie Callan}.} \bibinfo{year}{2023}\natexlab{}.
\newblock \showarticletitle{Precise Zero-Shot Dense Retrieval without Relevance Labels}. In \bibinfo{booktitle}{\emph{Proceedings of the 61st Annual Meeting of the Association for Computational Linguistics (Volume 1: Long Papers)}}. \bibinfo{pages}{1762--1777}.
\newblock


\bibitem[\protect\citeauthoryear{Gardner, Talukdar, Krishnamurthy, and Mitchell}{Gardner et~al\mbox{.}}{2014}]%
        {kgvector}
\bibfield{author}{\bibinfo{person}{Matt Gardner}, \bibinfo{person}{Partha~Pratim Talukdar}, \bibinfo{person}{Jayant Krishnamurthy}, {and} \bibinfo{person}{Tom~M. Mitchell}.} \bibinfo{year}{2014}\natexlab{}.
\newblock \showarticletitle{Incorporating Vector Space Similarity in Random Walk Inference over Knowledge Bases}. In \bibinfo{booktitle}{\emph{EMNLP}}. \bibinfo{pages}{397--406}.
\newblock


\bibitem[\protect\citeauthoryear{Guu, Lee, Tung, Pasupat, and Chang}{Guu et~al\mbox{.}}{2020}]%
        {rag_icml20}
\bibfield{author}{\bibinfo{person}{Kelvin Guu}, \bibinfo{person}{Kenton Lee}, \bibinfo{person}{Zora Tung}, \bibinfo{person}{Panupong Pasupat}, {and} \bibinfo{person}{Ming{-}Wei Chang}.} \bibinfo{year}{2020}\natexlab{}.
\newblock \showarticletitle{Retrieval Augmented Language Model Pre-Training}. In \bibinfo{booktitle}{\emph{Proceedings of the 37th International Conference on Machine Learning, {ICML} 2020, 13-18 July 2020, Virtual Event}}, Vol.~\bibinfo{volume}{119}. \bibinfo{pages}{3929--3938}.
\newblock


\bibitem[\protect\citeauthoryear{Haoyu~Han}{Haoyu~Han}{2025}]%
        {graphrag}
\bibfield{author}{\bibinfo{person}{Harry Shomer Kai Guo Jiayuan Ding Yongjia Lei Mahantesh Halappanavar Ryan A. Rossi Subhabrata Mukherjee Xianfeng Tang Qi He Zhigang Hua Bo Long Tong Zhao Neil Shah Amin Javari Yinglong Xia Jiliang~Tang Haoyu~Han, Yu~Wang}.} \bibinfo{year}{2025}\natexlab{}.
\newblock \showarticletitle{Retrieval-Augmented Generation with Graphs (GraphRAG)}.
\newblock \bibinfo{journal}{\emph{CoRR}}  \bibinfo{volume}{abs/2501.00309} (\bibinfo{year}{2025}).
\newblock


\bibitem[\protect\citeauthoryear{Herzig, M{\"{u}}ller, Krichene, and Eisenschlos}{Herzig et~al\mbox{.}}{2021}]%
        {sigtabret}
\bibfield{author}{\bibinfo{person}{Jonathan Herzig}, \bibinfo{person}{Thomas M{\"{u}}ller}, \bibinfo{person}{Syrine Krichene}, {and} \bibinfo{person}{Julian~Martin Eisenschlos}.} \bibinfo{year}{2021}\natexlab{}.
\newblock \showarticletitle{Open Domain Question Answering over Tables via Dense Retrieval}. In \bibinfo{booktitle}{\emph{NAACL-HLT}}. \bibinfo{pages}{512--519}.
\newblock


\bibitem[\protect\citeauthoryear{Hu, Shen, Wallis, Allen{-}Zhu, Li, Wang, Wang, and Chen}{Hu et~al\mbox{.}}{2022}]%
        {lora}
\bibfield{author}{\bibinfo{person}{Edward~J. Hu}, \bibinfo{person}{Yelong Shen}, \bibinfo{person}{Phillip Wallis}, \bibinfo{person}{Zeyuan Allen{-}Zhu}, \bibinfo{person}{Yuanzhi Li}, \bibinfo{person}{Shean Wang}, \bibinfo{person}{Lu Wang}, {and} \bibinfo{person}{Weizhu Chen}.} \bibinfo{year}{2022}\natexlab{}.
\newblock \showarticletitle{LoRA: Low-Rank Adaptation of Large Language Models}. In \bibinfo{booktitle}{\emph{The Tenth International Conference on Learning Representations, {ICLR} 2022, Virtual Event, April 25-29, 2022}}. \bibinfo{publisher}{OpenReview.net}.
\newblock
\urldef\tempurl%
\url{https://openreview.net/forum?id=nZeVKeeFYf9}
\showURL{%
\tempurl}


\bibitem[\protect\citeauthoryear{Hu, Bieker, Li, Jiang, Keigwin, Ranganath, Keutzer, and Upadhyay}{Hu et~al\mbox{.}}{2024}]%
        {routebench}
\bibfield{author}{\bibinfo{person}{Qitian~Jason Hu}, \bibinfo{person}{Jacob Bieker}, \bibinfo{person}{Xiuyu Li}, \bibinfo{person}{Nan Jiang}, \bibinfo{person}{Benjamin Keigwin}, \bibinfo{person}{Gaurav Ranganath}, \bibinfo{person}{Kurt Keutzer}, {and} \bibinfo{person}{Shriyash~Kaustubh Upadhyay}.} \bibinfo{year}{2024}\natexlab{}.
\newblock \showarticletitle{RouterBench: {A} Benchmark for Multi-LLM Routing System}.
\newblock \bibinfo{journal}{\emph{CoRR}}  \bibinfo{volume}{abs/2403.12031} (\bibinfo{year}{2024}).
\newblock


\bibitem[\protect\citeauthoryear{Jiang, Ren, and Lin}{Jiang et~al\mbox{.}}{2023a}]%
        {routellmreward}
\bibfield{author}{\bibinfo{person}{Dongfu Jiang}, \bibinfo{person}{Xiang Ren}, {and} \bibinfo{person}{Bill~Yuchen Lin}.} \bibinfo{year}{2023}\natexlab{a}.
\newblock \showarticletitle{LLM-Blender: Ensembling Large Language Models with Pairwise Ranking and Generative Fusion}. In \bibinfo{booktitle}{\emph{ACL}}. \bibinfo{pages}{14165--14178}.
\newblock


\bibitem[\protect\citeauthoryear{Jiang, Xu, Gao, Sun, Liu, Dwivedi{-}Yu, Yang, Callan, and Neubig}{Jiang et~al\mbox{.}}{2023b}]%
        {rag_chunk_23}
\bibfield{author}{\bibinfo{person}{Zhengbao Jiang}, \bibinfo{person}{Frank~F. Xu}, \bibinfo{person}{Luyu Gao}, \bibinfo{person}{Zhiqing Sun}, \bibinfo{person}{Qian Liu}, \bibinfo{person}{Jane Dwivedi{-}Yu}, \bibinfo{person}{Yiming Yang}, \bibinfo{person}{Jamie Callan}, {and} \bibinfo{person}{Graham Neubig}.} \bibinfo{year}{2023}\natexlab{b}.
\newblock \showarticletitle{Active Retrieval Augmented Generation}. In \bibinfo{booktitle}{\emph{EMNLP 2023}}. \bibinfo{pages}{7969--7992}.
\newblock


\bibitem[\protect\citeauthoryear{Kim, Kim, Lee, and Shin}{Kim et~al\mbox{.}}{2025}]%
        {dpo_kim2025spread}
\bibfield{author}{\bibinfo{person}{Dongyoung Kim}, \bibinfo{person}{Jaehyung Kim}, \bibinfo{person}{Kimin Lee}, {and} \bibinfo{person}{Jinwoo Shin}.} \bibinfo{year}{2025}\natexlab{}.
\newblock \showarticletitle{Spread Preference Annotation: Direct Preference Judgment for Efficient {LLM} Alignment}. In \bibinfo{booktitle}{\emph{The Thirteenth International Conference on Learning Representations}}.
\newblock
\urldef\tempurl%
\url{https://openreview.net/forum?id=BPgK5XW1Nb}
\showURL{%
\tempurl}


\bibitem[\protect\citeauthoryear{Leis, Gubichev, Mirchev, Boncz, Kemper, and Neumann}{Leis et~al\mbox{.}}{2015}]%
        {card}
\bibfield{author}{\bibinfo{person}{Viktor Leis}, \bibinfo{person}{Andrey Gubichev}, \bibinfo{person}{Atanas Mirchev}, \bibinfo{person}{Peter~A. Boncz}, \bibinfo{person}{Alfons Kemper}, {and} \bibinfo{person}{Thomas Neumann}.} \bibinfo{year}{2015}\natexlab{}.
\newblock \showarticletitle{How Good Are Query Optimizers, Really?}
\newblock \bibinfo{journal}{\emph{Proc. {VLDB} Endow.}} \bibinfo{volume}{9}, \bibinfo{number}{3} (\bibinfo{year}{2015}), \bibinfo{pages}{204--215}.
\newblock


\bibitem[\protect\citeauthoryear{Li, Luo, Chai, Li, and Tang}{Li et~al\mbox{.}}{2024}]%
        {NLP2SQLDawn}
\bibfield{author}{\bibinfo{person}{Boyan Li}, \bibinfo{person}{Yuyu Luo}, \bibinfo{person}{Chengliang Chai}, \bibinfo{person}{Guoliang Li}, {and} \bibinfo{person}{Nan Tang}.} \bibinfo{year}{2024}\natexlab{}.
\newblock \showarticletitle{The Dawn of Natural Language to {SQL:} Are We Fully Ready? [Experiment, Analysis {\&} Benchmark {]}}.
\newblock \bibinfo{journal}{\emph{Proc. {VLDB} Endow.}} \bibinfo{volume}{17}, \bibinfo{number}{11} (\bibinfo{year}{2024}), \bibinfo{pages}{3318--3331}.
\newblock


\bibitem[\protect\citeauthoryear{Loshchilov and Hutter}{Loshchilov and Hutter}{2019}]%
        {loshchilov2017decoupled}
\bibfield{author}{\bibinfo{person}{Ilya Loshchilov} {and} \bibinfo{person}{Frank Hutter}.} \bibinfo{year}{2019}\natexlab{}.
\newblock \showarticletitle{Decoupled Weight Decay Regularization}. In \bibinfo{booktitle}{\emph{7th International Conference on Learning Representations, {ICLR} 2019, New Orleans, LA, USA, May 6-9, 2019}}.
\newblock


\bibitem[\protect\citeauthoryear{Ma, Xu, Jiang, Li, Qu, and Guo}{Ma et~al\mbox{.}}{2024}]%
        {tog2}
\bibfield{author}{\bibinfo{person}{Shengjie Ma}, \bibinfo{person}{Chengjin Xu}, \bibinfo{person}{Xuhui Jiang}, \bibinfo{person}{Muzhi Li}, \bibinfo{person}{Huaren Qu}, {and} \bibinfo{person}{Jian Guo}.} \bibinfo{year}{2024}\natexlab{}.
\newblock \showarticletitle{Think-on-Graph 2.0: Deep and Interpretable Large Language Model Reasoning with Knowledge Graph-guided Retrieval}.
\newblock \bibinfo{journal}{\emph{CoRR}}  \bibinfo{volume}{abs/2407.10805} (\bibinfo{year}{2024}).
\newblock


\bibitem[\protect\citeauthoryear{Ong, Almahairi, Wu, Chiang, Wu, Gonzalez, Kadous, and Stoica}{Ong et~al\mbox{.}}{2024}]%
        {routellm}
\bibfield{author}{\bibinfo{person}{Isaac Ong}, \bibinfo{person}{Amjad Almahairi}, \bibinfo{person}{Vincent Wu}, \bibinfo{person}{Wei{-}Lin Chiang}, \bibinfo{person}{Tianhao Wu}, \bibinfo{person}{Joseph~E. Gonzalez}, \bibinfo{person}{M.~Waleed Kadous}, {and} \bibinfo{person}{Ion Stoica}.} \bibinfo{year}{2024}\natexlab{}.
\newblock \showarticletitle{RouteLLM: Learning to Route LLMs with Preference Data}.
\newblock \bibinfo{journal}{\emph{CoRR}}  \bibinfo{volume}{abs/2406.18665} (\bibinfo{year}{2024}).
\newblock


\bibitem[\protect\citeauthoryear{Rafailov, Sharma, Mitchell, Manning, Ermon, and Finn}{Rafailov et~al\mbox{.}}{2023}]%
        {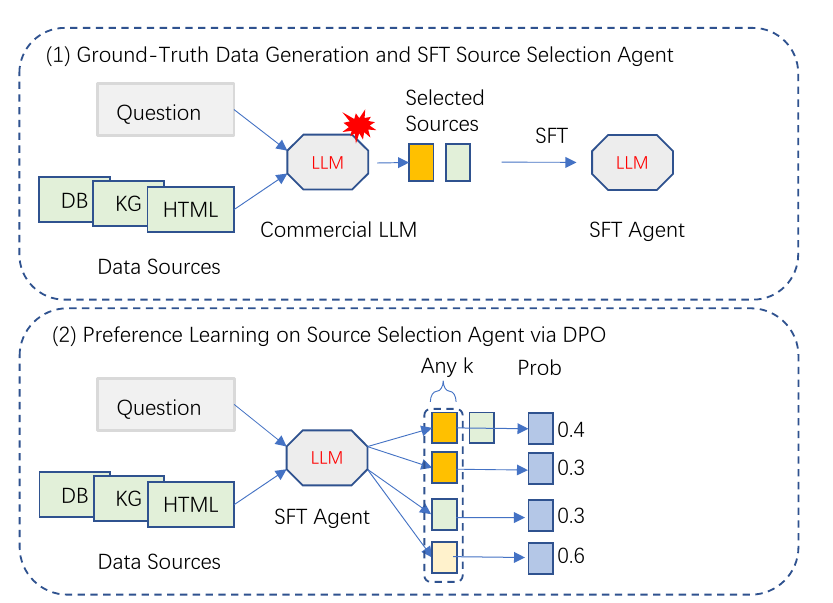}
\bibfield{author}{\bibinfo{person}{Rafael Rafailov}, \bibinfo{person}{Archit Sharma}, \bibinfo{person}{Eric Mitchell}, \bibinfo{person}{Christopher~D. Manning}, \bibinfo{person}{Stefano Ermon}, {and} \bibinfo{person}{Chelsea Finn}.} \bibinfo{year}{2023}\natexlab{}.
\newblock \showarticletitle{Direct Preference Optimization: Your Language Model is Secretly a Reward Model}. In \bibinfo{booktitle}{\emph{NeurIPS 2023}}.
\newblock


\bibitem[\protect\citeauthoryear{Robertson, Zaragoza, and Taylor}{Robertson et~al\mbox{.}}{2004}]%
        {bm25}
\bibfield{author}{\bibinfo{person}{Stephen~E. Robertson}, \bibinfo{person}{Hugo Zaragoza}, {and} \bibinfo{person}{Michael~J. Taylor}.} \bibinfo{year}{2004}\natexlab{}.
\newblock \showarticletitle{Simple {BM25} extension to multiple weighted fields}. In \bibinfo{booktitle}{\emph{CIKM 2004}}. \bibinfo{pages}{42--49}.
\newblock


\bibitem[\protect\citeauthoryear{Sarmah, Hall, Rao, Patel, Pasquali, and Mehta}{Sarmah et~al\mbox{.}}{2024}]%
        {hybridrag}
\bibfield{author}{\bibinfo{person}{Bhaskarjit Sarmah}, \bibinfo{person}{Benika Hall}, \bibinfo{person}{Rohan Rao}, \bibinfo{person}{Sunil Patel}, \bibinfo{person}{Stefano Pasquali}, {and} \bibinfo{person}{Dhagash Mehta}.} \bibinfo{year}{2024}\natexlab{}.
\newblock \showarticletitle{HybridRAG: Integrating Knowledge Graphs and Vector Retrieval Augmented Generation for Efficient Information Extraction}.
\newblock \bibinfo{journal}{\emph{CoRR}}  \bibinfo{volume}{abs/2408.04948} (\bibinfo{year}{2024}).
\newblock


\bibitem[\protect\citeauthoryear{Schulman, Wolski, Dhariwal, Radford, and Klimov}{Schulman et~al\mbox{.}}{2017}]%
        {ppo}
\bibfield{author}{\bibinfo{person}{John Schulman}, \bibinfo{person}{Filip Wolski}, \bibinfo{person}{Prafulla Dhariwal}, \bibinfo{person}{Alec Radford}, {and} \bibinfo{person}{Oleg Klimov}.} \bibinfo{year}{2017}\natexlab{}.
\newblock \showarticletitle{Proximal Policy Optimization Algorithms}.
\newblock \bibinfo{journal}{\emph{CoRR}}  \bibinfo{volume}{abs/1707.06347} (\bibinfo{year}{2017}).
\newblock


\bibitem[\protect\citeauthoryear{Sevgili, Shelmanov, Arkhipov, Panchenko, and Biemann}{Sevgili et~al\mbox{.}}{2022}]%
        {sevgili2022neural}
\bibfield{author}{\bibinfo{person}{{\"O}zge Sevgili}, \bibinfo{person}{Artem Shelmanov}, \bibinfo{person}{Mikhail Arkhipov}, \bibinfo{person}{Alexander Panchenko}, {and} \bibinfo{person}{Chris Biemann}.} \bibinfo{year}{2022}\natexlab{}.
\newblock \showarticletitle{Neural entity linking: A survey of models based on deep learning}.
\newblock \bibinfo{journal}{\emph{Semantic Web}} \bibinfo{volume}{13}, \bibinfo{number}{3} (\bibinfo{year}{2022}), \bibinfo{pages}{527--570}.
\newblock


\bibitem[\protect\citeauthoryear{Sugiyama, Hatano, Yoshikawa, and Uemura}{Sugiyama et~al\mbox{.}}{2003}]%
        {tfidf}
\bibfield{author}{\bibinfo{person}{Kazunari Sugiyama}, \bibinfo{person}{Kenji Hatano}, \bibinfo{person}{Masatoshi Yoshikawa}, {and} \bibinfo{person}{Shunsuke Uemura}.} \bibinfo{year}{2003}\natexlab{}.
\newblock \showarticletitle{Refinement of {TF-IDF} schemes for web pages using their hyperlinked neighboring pages}. In \bibinfo{booktitle}{\emph{HYPERTEXT 2003}}. \bibinfo{pages}{198--207}.
\newblock


\bibitem[\protect\citeauthoryear{Sun, Xu, Tang, Wang, Lin, Gong, Ni, Shum, and Guo}{Sun et~al\mbox{.}}{2024}]%
        {tog1}
\bibfield{author}{\bibinfo{person}{Jiashuo Sun}, \bibinfo{person}{Chengjin Xu}, \bibinfo{person}{Lumingyuan Tang}, \bibinfo{person}{Saizhuo Wang}, \bibinfo{person}{Chen Lin}, \bibinfo{person}{Yeyun Gong}, \bibinfo{person}{Lionel~M. Ni}, \bibinfo{person}{Heung{-}Yeung Shum}, {and} \bibinfo{person}{Jian Guo}.} \bibinfo{year}{2024}\natexlab{}.
\newblock \showarticletitle{Think-on-Graph: Deep and Responsible Reasoning of Large Language Model on Knowledge Graph}. In \bibinfo{booktitle}{\emph{ICLR}}.
\newblock


\bibitem[\protect\citeauthoryear{Unger, B{\"{u}}hmann, Lehmann, Ngomo, Gerber, and Cimiano}{Unger et~al\mbox{.}}{2012}]%
        {templateparsekg}
\bibfield{author}{\bibinfo{person}{Christina Unger}, \bibinfo{person}{Lorenz B{\"{u}}hmann}, \bibinfo{person}{Jens Lehmann}, \bibinfo{person}{Axel{-}Cyrille~Ngonga Ngomo}, \bibinfo{person}{Daniel Gerber}, {and} \bibinfo{person}{Philipp Cimiano}.} \bibinfo{year}{2012}\natexlab{}.
\newblock \showarticletitle{Template-based question answering over {RDF} data}. In \bibinfo{booktitle}{\emph{WWW}}. \bibinfo{pages}{639--648}.
\newblock


\bibitem[\protect\citeauthoryear{Wolf, Debut, Sanh, Chaumond, Delangue, Moi, Cistac, Rault, Louf, Funtowicz, et~al\mbox{.}}{Wolf et~al\mbox{.}}{2019}]%
        {wolf2019huggingface}
\bibfield{author}{\bibinfo{person}{Thomas Wolf}, \bibinfo{person}{Lysandre Debut}, \bibinfo{person}{Victor Sanh}, \bibinfo{person}{Julien Chaumond}, \bibinfo{person}{Clement Delangue}, \bibinfo{person}{Anthony Moi}, \bibinfo{person}{Pierric Cistac}, \bibinfo{person}{Tim Rault}, \bibinfo{person}{R{\'e}mi Louf}, \bibinfo{person}{Morgan Funtowicz}, {et~al\mbox{.}}} \bibinfo{year}{2019}\natexlab{}.
\newblock \showarticletitle{Huggingface's transformers: State-of-the-art natural language processing}.
\newblock \bibinfo{journal}{\emph{arXiv preprint arXiv:1910.03771}} (\bibinfo{year}{2019}).
\newblock


\bibitem[\protect\citeauthoryear{Xia, Chen, and Gao}{Xia et~al\mbox{.}}{2024}]%
        {xia2024}
\bibfield{author}{\bibinfo{person}{Yikuan Xia}, \bibinfo{person}{Jiazun Chen}, {and} \bibinfo{person}{Jun Gao}.} \bibinfo{year}{2024}\natexlab{}.
\newblock \showarticletitle{Winning Solution For Meta KDD Cup' 24}.
\newblock \bibinfo{journal}{\emph{CoRR}}  \bibinfo{volume}{abs/2410.00005} (\bibinfo{year}{2024}).
\newblock


\bibitem[\protect\citeauthoryear{Xiao~Yang}{Xiao~Yang}{2024}]%
        {crag2024}
\bibfield{author}{\bibinfo{person}{Hao Xin Yushi Sun Nikita Bhalla Xiangsen Chen Sajal Choudhary Rongze Daniel Gui Ziran Will Jiang Ziyu Jiang Lingkun Kong Brian Moran Jiaqi Wang Yifan Ethan Xu An Yan Chenyu Yang Eting Yuan Hanwen Zha Nan Tang Lei Chen Nicolas Scheffer Yue Liu Nirav Shah Rakesh Wanga Anuj Kumar Wen-tau Yih Xin Luna~Dong Xiao~Yang, Kai~Sun}.} \bibinfo{year}{2024}\natexlab{}.
\newblock \showarticletitle{CRAG - Comprehensive RAG Benchmark}.
\newblock \bibinfo{journal}{\emph{CoRR}}  \bibinfo{volume}{abs/2406.04744} (\bibinfo{year}{2024}).
\newblock


\bibitem[\protect\citeauthoryear{Yang, Sun, Xin, Sun, Bhalla, Chen, Choudhary, Gui, Jiang, Jiang, et~al\mbox{.}}{Yang et~al\mbox{.}}{2024}]%
        {yang2024crag}
\bibfield{author}{\bibinfo{person}{Xiao Yang}, \bibinfo{person}{Kai Sun}, \bibinfo{person}{Hao Xin}, \bibinfo{person}{Yushi Sun}, \bibinfo{person}{Nikita Bhalla}, \bibinfo{person}{Xiangsen Chen}, \bibinfo{person}{Sajal Choudhary}, \bibinfo{person}{Rongze~Daniel Gui}, \bibinfo{person}{Ziran~Will Jiang}, \bibinfo{person}{Ziyu Jiang}, {et~al\mbox{.}}} \bibinfo{year}{2024}\natexlab{}.
\newblock \showarticletitle{CRAG--Comprehensive RAG Benchmark}.
\newblock \bibinfo{journal}{\emph{arXiv preprint arXiv:2406.04744}} (\bibinfo{year}{2024}).
\newblock


\bibitem[\protect\citeauthoryear{Yao, Gu, Cong, Jin, and Lv}{Yao et~al\mbox{.}}{2022}]%
        {entitygraph}
\bibfield{author}{\bibinfo{person}{Dezhong Yao}, \bibinfo{person}{Yuhong Gu}, \bibinfo{person}{Gao Cong}, \bibinfo{person}{Hai Jin}, {and} \bibinfo{person}{Xinqiao Lv}.} \bibinfo{year}{2022}\natexlab{}.
\newblock \showarticletitle{Entity Resolution with Hierarchical Graph Attention Networks}. In \bibinfo{booktitle}{\emph{SIGMOD}}. \bibinfo{pages}{429--442}.
\newblock


\bibitem[\protect\citeauthoryear{Yih, Chang, He, and Gao}{Yih et~al\mbox{.}}{2015}]%
        {sematicparsekg}
\bibfield{author}{\bibinfo{person}{Wen{-}tau Yih}, \bibinfo{person}{Ming{-}Wei Chang}, \bibinfo{person}{Xiaodong He}, {and} \bibinfo{person}{Jianfeng Gao}.} \bibinfo{year}{2015}\natexlab{}.
\newblock \showarticletitle{Semantic Parsing via Staged Query Graph Generation: Question Answering with Knowledge Base}. In \bibinfo{booktitle}{\emph{ACL}}. \bibinfo{pages}{1321--1331}.
\newblock


\end{thebibliography}
 

\end{document}